\documentclass[english, pra, twocolumn,superscriptaddress,secnumarabic,aps,nobibnotes,numerical,floatfix]{revtex4-1}

\usepackage{tabularx}
\usepackage{enumerate}
\usepackage[11pt]{moresize}
\usepackage[normalem]{ulem}
\usepackage{amssymb}
\usepackage{xcolor}
\usepackage{amsmath}
\usepackage{float}
\usepackage{amstext}
\usepackage{graphicx}
\usepackage{esint}
\usepackage{color}
\usepackage{physics}
\usepackage[colorlinks]{hyperref}
\usepackage{amsthm}
\usepackage{times}
\usepackage{mathrsfs }
\usepackage[toc,page]{appendix}
\usepackage{booktabs}
\usepackage{mathtools}
\usepackage[caption=false]{subfig}
\definecolor{mygreen}{RGB}{0, 122, 100}

\usepackage[colorlinks]{hyperref}

\newcommand{\smx}{\sigma^x}
\newcommand{\smy}{\sigma^y}
\newcommand{\smz}{\sigma^z}
\newcommand{\smp}{\sigma^+}
\newcommand{\smm}{\sigma^-}

\newcommand{\half}{\frac{1}{2}}
\newcommand{\od}{\omega^{\rm d}}
\newcommand{\hc}{\text{h.c.}}

\newcommand{\Heff}{H_{\rm eff}}

\newcommand{\ophi}{\omega^\phi}

\newcommand{\otb}{\omega_{\rm\small c}}

\begin{document}
	\title{Adiabatic quantum simulations with driven superconducting qubits}
	\author{Marco Roth}
	\affiliation{JARA Institute for Quantum Information (PGI-11), Forschungszentrum J\"ulich, 52428 J\"ulich, Germany}
	\affiliation{Institute for Quantum Information, RWTH Aachen University, 52056 Aachen, Germany}
	\affiliation{IBM Research -- Z\"urich, 8803 R\"uschlikon, Switzerland}
	\author{Nikolaj Moll}
	\author{Gian Salis}
	\author{Marc Ganzhorn}
	\author{Daniel J. Egger}
	\author{Stefan Filipp}
	\affiliation{IBM Research -- Z\"urich, 8803 R\"uschlikon, Switzerland}
	\author{Sebastian Schmidt}
	\affiliation{Institute for Theoretical Physics, ETH Z\"urich, 8093 Zurich, Switzerland}
	\date{\today}
	\begin{abstract}
	We propose a quantum simulator based on driven superconducting qubits where the interactions are generated parametrically by a polychromatic magnetic flux modulation of a tunable bus element. Using a time-dependent Schrieffer-Wolff transformation, we analytically derive a multi-qubit Hamiltonian which features independently tunable $XX$ and $YY$-type interactions as well as local bias fields over a large parameter range. We demonstrate the adiabatic simulation of the ground state of a hydrogen molecule using two superconducting qubits and one tunable bus element. The time required to reach chemical accuracy lies in the few microsecond range and therefore could be implemented on currently available superconducting circuits. Further applications of this technique may also be found in the simulation of interacting spin systems.
	\end{abstract}
	\maketitle
\section{Introduction}
The simulation of a quantum system with a classical computer is notoriously difficult as the number of parameters needed to describe a quantum state grows exponentially with increasing system size. It has been recognized early that these scaling issues can be circumvented by using a controllable quantum system as a simulator~\cite{Feynman1982}. Quantum simulations \cite{Buluta2009, Georgescu2014} have been successfully implemented in NMR \cite{Du2010, Lu2011}, photonic devices \cite{Lanyon2010, Aspuru-Guzik2012, Peruzzo2014, Hartmann2016}, ultracold atoms \cite{Bloch2012, Schaetz2013}, trapped ions \cite{Blatt2012, Zhang2017} and superconducting qubits \cite{Schmidt2013, Salathe2015, Barends2016,  Langford2017, OMalley2016, Roushan2016, Wendin2017}. These experiments mainly use gate based methods \cite{Heras2014, Salathe2015,  Barends2015}, in which the dynamics of the system is discretized into Trotterized time-steps \cite{Suzuki1990, Langford2017}, or static properties such as ground states are calculated using a variational quantum eigensolver \cite{OMalley2016, Kandala2017}. In contrast, analogue simulations that directly implement the desired Hamiltonian in hardware have been proposed \cite{Farhi2000, Ivan2011, Egger2013, Babbush2014, Kyriienko2018} and realized \cite{Barends2016, Roushan2016, Braumuller2017}. These simulations are typically based on the adiabatic theorem \cite{Schiff1955}. In such a protocol, the system is initialized in an easily accessible eigenstate. Subsequently, the Hamiltonian parameters are varied slowly such that the Hamiltonian at the end of the evolution is equivalent to some target Hamiltonian. If the change is performed sufficiently slow, the state of the system will remain an eigenstate throughout the evolution and adiabatically follow the changing Hamiltonian. The advantages of this approach are for example that the qubits can be tuned simultaneously to efficiently generate entangled eigenstates of the target Hamiltonian and that there is no need for Trotterization.

One particular interesting application of quantum simulations is quantum chemistry. Here, the fermionic degrees of freedom are mapped to qubit operators \cite{Jordan1928, Bravyi2002}. In general, these mappings create $k$-local terms that are challenging to implement in experiments. To circumvent this problem, several methods such as perturbative gadgets \cite{Babbush2014, Cao2015} or alternative mapping schemes \cite{Lechnere1500838} have been employed to reduce the problem to Hamiltonians that only feature two-local interactions. In this paper, we focus on a particular subset of two-local Hamiltonians
 	\begin{equation}
 	H=\sum_{i}^N \left(a_i\smz_i + b_i\smx_i + c_i\smy_i\right) + \sum_{i<j}\left(J^x_{ij} \smx_i\smx_j + J^y_{ij} \smy_i\smy_j \right)\, ,\label{eq:generic_H}
 	\end{equation}
where $a_i$, $b_i$ and $c_i$ are the coefficients of the single qubit terms and $J^x_{ij}$ and $J^y_{ij}$ are the coefficients of the two-qubit $XX$- and $YY$-interactions, respectively. Such two-qubit interactions can be realized with superconducting qubits using parametric coupling schemes \cite{McKay2016, Roth2017, Didier2018, Reagoreaao3603, Roushan2016, doi:10.1063/1.4919759}. The interactions are mediated by harmonically modulating the frequency of a tunable coupler. The possibility of using fixed-frequency qubits is among the benefits of this method. This reduces the sensitivity to flux noise which is reflected in increased coherence times compared to approaches that require frequency tunabe qubits \cite{DiCarlo2009, Salathe2015}. Such a parametric setup can generate many different types of interactions such as the iSWAP-type interaction that couples the $\ket{10}$ and $\ket{01}$ states and the bSWAP-type interaction which induces transitions between the  $\ket{00}$ and $\ket{11}$ states \cite{Roth2017}. In this paper we, propose to realize independently tunable $XX$ and $YY$-type interactions by combining iSWAP and bSWAP-type interactions using a bichromatic flux pulse similar to the qubit-resonator entanglement scheme in \cite{Lu2017}.\\
Analytic expressions for these effective interactions are obtained by eliminating the coupler degrees of freedom with a generalized Schrieffer-Wolff transformation \cite{Roth2017, Theis2017} that accounts for the dynamics of the coupler. By explicitly considering the counter-rotating terms, we are able to accurately describe the effective two-qubit interactions as shown by a comparison to numerical simulations. This method has been previously used in an undriven system with two qubits and a monochromatically modulated tunable coupler \cite{Roth2017}. Here, we generalize this approach to $N$ qubits interacting with one coupler modulated at $M$ frequencies. Furthermore, we show the applicability of this approach in the weak-driving limit.\\ 
As an example of a quantum simulation we adiabatically determine the ground state of the hydrogen molecule. The single qubit terms of the Hamiltonian are realized by driving the qubits via a coherent microwave tone. In the rotating frame of the drive, arbitrary longitudinal fields as well as transverse fields in a wide parameter range can be realized. Using a standard Markovian master equation, we show that the external driving results in a non-equilibrium steady state that attains a maximum fidelity at an optimal protocol run time. For typical coherence parameters, this optimum is reached after a few microseconds.
 \begin{figure}[tb]
 	\centering
 	\includegraphics[width=0.3\paperwidth]{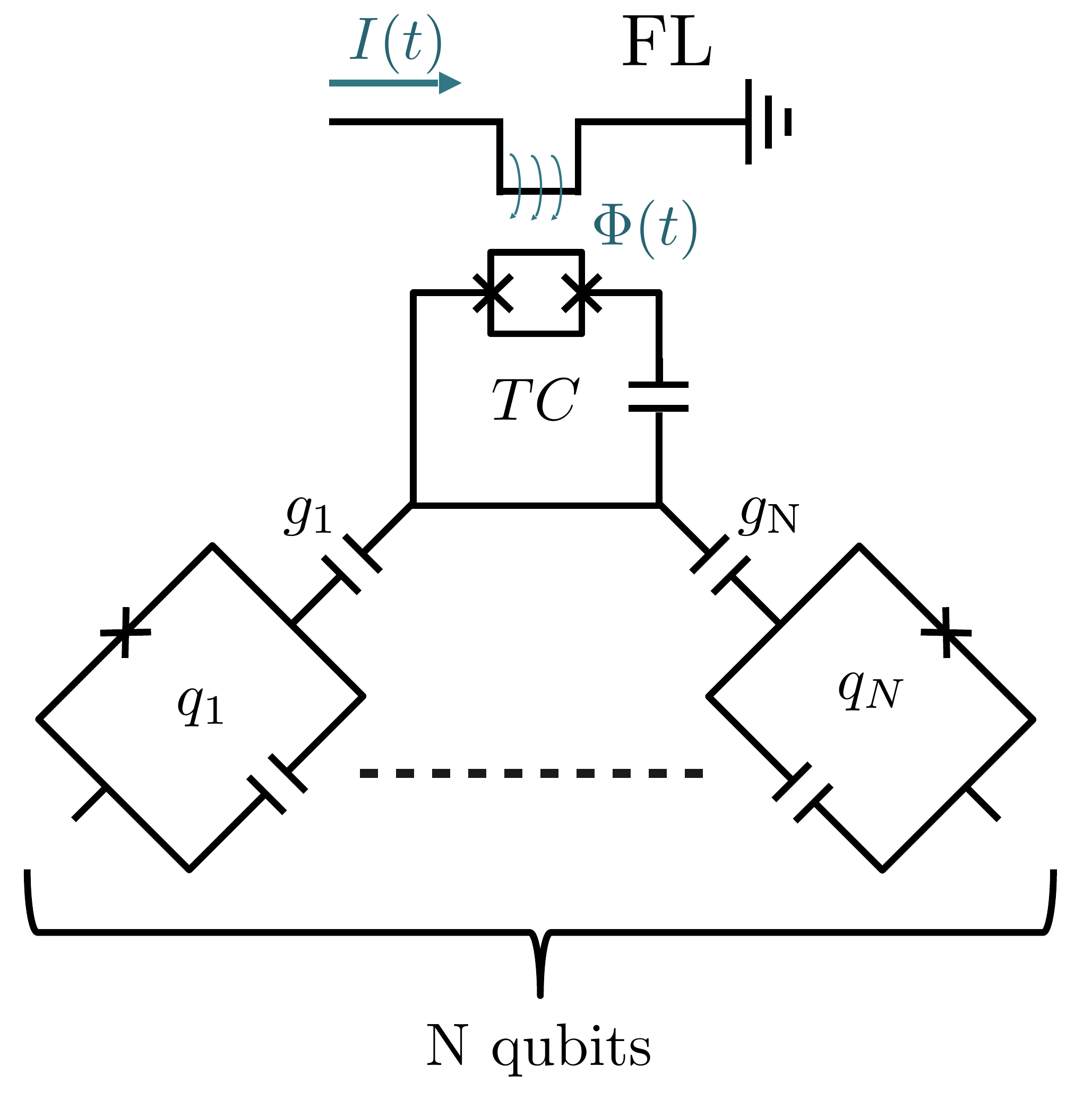}
 	\caption{Schematic of $N$ qubits $q_1,\dots,q_N$ with fixed frequencies $\omega_1\,\dots,\omega_{\rm N}$ coupled to a tunable coupler (TC) with coupling strenghts $g_1,\dots, g_{\rm N}$. The frequency of the coupler can be controlled via a current $I(t)$ through a flux line (FL) resulting in a flux $\Phi(t)$ threading the superconducting quantum interference device (SQUID) loop.}
 	\label{fig:figure1}
 \end{figure}
The paper is structured in the following way. Section~\ref{sec:model} discusses the general setup and introduces the physical model. The effective Hamiltonian of the superconducting qubit device is derived in Sections~\ref{sec:dispersive_regime} and~\ref{sec:effective_hamiltonian}. Section ~\ref{sec:ising_interactions} shows a numerical validation of the two-qubit interactions obtained by a bichromatic modulation of the coupler. Finally, Section~\ref{sec:adiabatic_annealing} presents numerical results of the adiabatic simulation of the hydrogen molecule which is followed by the conclusion in section~\ref{sec:conclusion}.

\section{Model}
\label{sec:model}
We study a quantum device with $N$ computational qubits with frequencies $\omega_j$ ($j = 1, ..., N$) transversely coupled to a tunable coupler with time-dependent frequency $\omega_{\rm c}(t)$.
A schematic of the setup using transmon qubits \cite{Koch2007} is shown in Fig.~\ref{fig:figure1}. The Hamiltonian is given by \cite{McKay2016, Roth2017}
\begin{align}
H_{\rm Tr}(t) &= \sum_{j}^N \Big[\omega_j a^\dag_ja_j + \frac{u_j}{2}a^\dag_ja_j(a^\dag_ja_j-1)\nonumber\\ 
&+\frac{f_j}{2}\left(a^\dag_je^{-i(\od_j t+\varphi_j)}+a_je^{i(\od_j t+\varphi_j)}\right)\Big]\nonumber\\
& + \omega_{\rm c}(t) a^\dag_{\rm c}a_{\rm c} + \frac{u_{\rm c}}{2}a^\dag_{\rm c}a_{\rm c}(a^\dag_{\rm c}a_{\rm c}-1)\nonumber\\
& + \sum_{j}^N g_j(a_j^\dag + a_j)(a_{\rm c}^\dag + a_{\rm c}) \label{eq:full_transmon_H} \,,
\end{align}
where $a_j$ ($a_j^\dag$) and $a_{\rm c}$ ($a_{\rm c}^\dag$) are bosonic annihilation (creation) operators. Here, we use the convention $\hbar=1$. The computational qubits are subject to an external drive of strength $f_j$, frequency $\od_j$ and phase $\varphi_j$. The coupling strength between the  computational qubits and the tunable coupler is $g_j$. Each transmon has an anharmonicity  $u_j$. The frequency of the tunable coupler in Eq.~(\ref{eq:full_transmon_H}) can be tuned with an external flux parameter $\Phi(t)$ according to
\begin{equation}
\label{eq:couplerfrequency}
\omega_{\rm c}(t)=\omega_{\rm c}^{0}\sqrt{\abs{\cos(\pi \Phi(t) /\phi_0)}} \, ,
\end{equation}
where $\phi_0$ is the flux quantum and $\omega_{\rm c}^{\rm 0}$ is the coupler frequency at zero flux. For large anharmonicities, the bosonic operators can be restricted to a two-level state space consisting of the ground state and the first excited state of the transmon. In this two-level approximation the Hamiltonian in Eq.~(\ref{eq:full_transmon_H}) reduces to 
\begin{align}
H(t) = \sum_{j}^N&\left[ -\frac{\omega_j}{2}\sigma^z_j +  \frac{f_j}{2}\left(e^{-i \left( \od_j t + \varphi_j \right)} \smp_j  + {\rm h.c.} \right) \right] \nonumber\\
&-\frac{\omega_{\rm c}(t)}{2}\sigma^z_{\rm c} + \sum_j g_j\sigma^x_j\sigma^x_{\rm c} \,. \label{eq:h_three_qubits}
\end{align}
Here, $\sigma_j^\alpha$ and $\sigma_{\rm c}^\alpha$ ($\alpha=x,y,z$) denote the standard Pauli operators and $\sigma^\pm_j=\left(\smx_j\pm i\smy_j\right)/2$ is the raising (lowering) operator of qubit $j$. We now consider a flux modulation 
\begin{align}
\Phi(t) =\theta +\sum_{m}^M \delta_m \cos(\ophi_m t)\,,\label{eq:phi_bi}
\end{align}
with dc-bias $\theta$ and $M$ independent harmonic tones of strength $\delta_m$ and frequency $\ophi_m$ \cite{McKay2016, Roth2017}.
By expanding the frequency of the coupler in Eq.~(\ref{eq:couplerfrequency}) in powers of the modulation amplitudes $\delta_m$, we obtain harmonic modulations of the coupler frequency
\begin{align}
\otb(t) & \approx \omega_{\rm c}^\theta + \partial_\Phi\otb \big|_{\Phi=\theta} \sum_m^M  \delta_m \cos(\ophi_m t) + \mathcal{O}(\delta_m^2/\phi_0^2) \, ,
\label{eq:otbexpand_bichromatic}
\end{align}
with $\omega_{\rm c}^\theta=\omega_{\rm c}^0\sqrt{\abs{\cos(\pi\theta/\phi_0)}}$. A monochromatic flux modulation with a single tone ($M = 1$) allows for the activation of iSWAP ($\sim XX+YY$) or bSWAP-type ($\sim XX-YY$) interactions between the computational qubits \cite{Roth2017}. 
In the following, we show that a bichromatic modulation of the coupler ($M = 2$) combined with external driving of the computational qubits allows us to simulate independent $XX$ and $YY$-type interactions and transverse fields.

\subsection{Dispersive regime in the rotating frame}
\label{sec:dispersive_regime}
We require the coupler to be detuned from both computational qubits such that $\abs{\omega_j-\omega_{\rm c}(t)} \gg g_j$ for all times $t$. For that purpose, we choose the modulation strengths $\delta_m$ in Eq.~(\ref{eq:phi_bi}) sufficiently small during the whole time evolution and the dc-bias $\theta$ such that $\abs{\Delta^\theta_{j,-}} \gg g_j $ with the qubit-coupler detunings $\Delta^\theta_{j,\pm} = \omega_j \pm \omega^\theta_c $. In this dispersive regime, the coupler degree of freedom is eliminated using a generalized Schrieffer-Wolff transformation (SWT) $U(t)=\exp{S(t)}$, where $S(t)$ is the anti-hermitian matrix
\begin{align}
S(t) =\sum_{j}^N\left(\alpha_{j,-}(t)\smp_j \sigma^-_{\rm c} + \alpha_{j,+}(t)\smp_j \sigma^+_{\rm c}- \hc \right)\,.
\label{eq:ansatz_S}
\end{align}
This transformation has been introduced in \cite{Roth2017} for the non-driven system. The parameters $\alpha_{j,\pm}(t)$ are chosen to be the solutions of the ordinary differential equation 
\begin{align}
i\dot{\alpha}_{j,\pm}(t) + g_j - \left[\omega_j\pm\omega_{\rm c}(t)\right] \alpha_{j,\pm}(t) &= 0\,.
\label{eq:deq_alpha}
\end{align}
With this choice, the computational subspace decouples from the tunable coupler resulting in an effective Hamiltonian where the coupler can be eliminated (see Appendix \ref{appendix_swt}). In Appendix~\ref{appendix_drive}, we argue that the external drive does not modify
this picture as long as the weak driving conditions is satisfied, i.e., $f_j \ll g_j$.

A second unitary transformation $R(t)=\exp(-i\sum_{j}\od_j\smz_j t/2)$ into a frame rotating at the drive frequencies then yields the time-dependent $N$-qubit Hamiltonian 
\begin{align}
\tilde{H}(t)= \sum_{j}^N&\left[\frac{\epsilon_j}{2} \sigma^z_j +   \frac{f_j}{2}\left( \cos(\varphi_j)\smx_j-\sin(\varphi_j)\smy_j\right)\right]\nonumber\\
+\sum_{i<j}&\left[\Omega^+_{ij}(t) e^{i \left(\omega_i^d - \omega^d_j \right) t} \smp_i\smm_j \right.\nonumber\\
+&\left.\Omega^-_{ij}(t) e^{i \left(\omega_i^d + \omega^d_j \right) t} \smp_i\smp_j +\hc\right]\,,\label{eq:effective_hamiltonian_rot}
\end{align}
with the drive-qubit detuning $\epsilon_j = \od_j-\bar{\omega}_j$. The dispersively shifted qubit frequencies $\bar{\omega}_j$ are given in Appendix~\ref{appendix_swt}. The last two lines in Eq.~(\ref{eq:effective_hamiltonian_rot}) describe iSWAP and bSWAP type interactions, respectively. The time-dependent coupling coefficients $\Omega^\pm_{ij}(t)$ can be decomposed into harmonics rotating at integer multiples of the modulation frequencies (see Appendix~\ref{app:alpha_fourier}), i.e.,
\begin{equation}
\Omega^{\pm}_{ij}(t) \approx \sum_{m}^M\sum_{k=-\infty}^\infty \bar{\Omega}^\pm_{ij}(k,m) e^{i k\ophi_m t}\,.\label{eq:omega_bichromatic}
\end{equation}
The amplitudes in Eq.~(\ref{eq:omega_bichromatic}) can be written explicitely as
\begin{align}
\label{eq:Omega_bar}
\bar{\Omega}^\pm_{ij}(k,m) =  \big[&g_i\bar{\alpha}_{j,-}(\mp k,m)+g_j\bar{\alpha}_{i,-}(k,m)\nonumber\\
 						         -&g_i\bar{\alpha}_{j,+}(\mp k,m)-g_j\bar{\alpha}_{i,+}(k,m)\big]/2\,,
\end{align}
where $\bar{\alpha}_{j,\pm}(k,m)$ are the Fourier components of the the $\alpha_{j,\pm}(t)$ parameters found in Eq.~(\ref{eq:deq_alpha}), i.e., 
\begin{equation}
\alpha_{j,\pm}(t) \approx\bar{\alpha}_{j,\pm}(0)+\sum_{m}^M\sum_{k\neq 0} \overline{\alpha}_{j,\pm}(k,m) e^{ik\ophi_m t}\,.
\end{equation}
In Appendix~\ref{app:alpha_fourier} we derive the following analytic expression for the case of bichromatic driving ($M=2$) valid to first order in the modulation strength, i.e.,
\begin{align}
\bar{\alpha}_{j,\pm}(k,m) &=g_j\sum_{q,p} \frac{J_{k-q}\left(\mp\lambda_m\right)J_q\left(\pm\lambda_m\right)J_{p}\left(\pm\lambda_{n}\right)^2}{q\ophi_m+p\ophi_{n}+\Delta^\theta_{j,\pm}}\,,\label{eq:alpha_bar}
\end{align}
with $n\neq m$ and
	\begin{equation}
	\bar{\alpha}_{j,\pm}(0) =g_j\sum_{q,p} 
	\frac{J_q\left(\pm\lambda_1\right)^2 J_{p}\left(\pm\lambda_{2}\right)^2}{q\ophi_1+p\ophi_{2}+\Delta^\theta_{j,\pm}}\,.
	\end{equation}
Here, $J_n(x)$ is the $n$-th Bessel function of the first kind. Furthermore, we have defined the effective modulation strength parameter $\lambda_m=\delta_m\partial_\Phi\otb \big|_{\Phi=\theta}/\ophi_m$ \cite{Roth2017}.

\subsection{Effective Hamiltonian}
\label{sec:effective_hamiltonian}
 In the following, we consider two qubits ($N = 2$) and a bichromatic modulation ($M = 2$). Assuming $\bar{\omega}_1 > \bar{\omega}_2$, static interactions can be obtained from Eq.~(\ref{eq:effective_hamiltonian_rot}) and Eq.~(\ref{eq:omega_bichromatic}) for modulation frequencies $\ophi_m = (\omega^d_i \pm \omega^d_j)/k$ with $-k\in\mathbb{N}$. The largest interaction strength corresponds to $|k| = 1$. We focus on the case of near resonant driving of the qubits $\od_j\approx\bar{\omega}_j$. Choosing the modulation frequencies to be resonant with the difference and sum of the qubit drive frequencies, i.e., $\ophi_{1} = \od_1 - \od_2$ and $\ophi_{2} = \od_1 + \od_2$, yields an effective, time-independent two-qubit Hamiltonian 
\begin{align}
\tilde{H}_{\rm eff} \approx \sum_{j=1,2} & 
\left[\frac{\epsilon_j}{2}\smz_j + \frac{f_j}{2}\left( \cos(\varphi_j)\smx_j-\sin(\varphi_j)\smy_j\right)\right]\nonumber\\ +&\frac{\Omega_{\rm x}}{2}  \smx_1\smx_2  +  \frac{\Omega_{\rm y}}{2} \smy_1\smy_2\,,\nonumber\\
\label{eq:effective_2qubithamiltonian}
\end{align}
where 
\begin{eqnarray}
\Omega_{\rm x}=\bar{\Omega}^+_{1,2}(-1,1) + \bar{\Omega}^{-}_{1,2}(-1,2)\,,\label{coupling_x}
\end{eqnarray} 
and 
\begin{eqnarray}
\Omega_{\rm y}=\bar{\Omega}^{+}_{1,2}(-1,1)-\bar{\Omega}^{-}_{1,2}(-1,2)\,.
\end{eqnarray} 
Here, only the term with $k = -1$ in Eq.~(\ref{eq:omega_bichromatic}) is relevant as the remaining terms are off-resonant and can be neglected as long as $\abs{g_1g_2/\Delta_j}\ll\omega_1-\omega_2$ for $j=1,2$. To first order in the modulation strength $\delta_m$, we obtain analytic expressions for these amplitudes (see Appendix~\ref{app:weakmodulation}), i.e.,
\begin{align}
\bar{\Omega}^{+}_{1,2}(-1,1) \approx &\delta_1\frac{g_1g_2}{2}\partial_\Phi\otb \big|_{\Phi=\theta}\nonumber\\\cross&\Big(\frac{1}{\Delta_{1,-}^\theta\Delta_{2,-}^\theta}+\frac{1}{\Delta_{1,+}^\theta\Delta_{2,+}^\theta}\Big)\,,\label{eq:eff_iswap}
\end{align}
and
\begin{align}
\bar{\Omega}^{-}_{1,2}(-1,2) \approx& -\delta_2\frac{g_1g_2}{2}\partial_\Phi\otb \big|_{\Phi=\theta}\nonumber\\\cross&\Big(\frac{1}{\Delta_{1,-}^\theta\Delta_{2,+}^\theta} +\frac{1}{\Delta_{1,+}^\theta\Delta_{2,-}^\theta}\Big)\,.\label{eq:eff_bswap}
\end{align}
The bichromatic modulation thus enables the independent tuning of $XX$- and $YY$-type interactions where the strength of the interaction mainly depends on the values of the modulation amplitudes $\delta_1$ and $\delta_2$ applied to the tunable coupler. At the same time, the external drive generates a transverse field which can be used to simulate the classical and the quantum Ising model as discussed in the next section. We note that the results in Eq.~(\ref{eq:eff_iswap}) and Eq.~(\ref{eq:eff_bswap}) are equivalent to the coupling strength of iSWAP and bSWAP gates as generated by a monochromatic modulation with $M = 1$ \cite{Roth2017}. However, already to second order in $\lambda_{1}$ and $\lambda_2$, the coupling strength of $XX$- and $YY$-type interactions cannot simply be obtained from a linear superposition of the respective monochromatic coupling strengths. This can be seen in Eq.~(\ref{eq:alpha_bar}), where the weight for the $m$-th modulation tone also depends on the strength and frequencies of all other tones. For the moment, we restrict ourselves to the case of resonant driving with the single-qubit drive frequency equal to the dispersively shifted qubit frequency, $\omega^d_j = \bar{\omega}_j$.  Fig.~\ref{fig:figure2} (a) shows a color plot of $\Omega_{\rm x}$ in the plane defined by the two modulation strengths. The corresponding values for $\Omega_{\rm y}$ can be seen in Fig.~\ref{fig:figure2} (b). The values for $\Omega_{\rm x}$ and $\Omega_{\rm y}$ are the result of pertubation theory to second order in $\delta_1$ and $\delta_2$. They are obtained by expanding the coupler frequency in Eq.~(\ref{eq:deq_alpha}) to second order in $\delta_1$ and $\delta_2$ and subsequent solution analogous to Appendix~\ref{app:alpha_fourier}.

\subsection{Special case: Ising interactions}
\label{sec:ising_interactions}
\begin{figure}[tb]
	\centering
	\includegraphics[width=0.35\paperwidth]{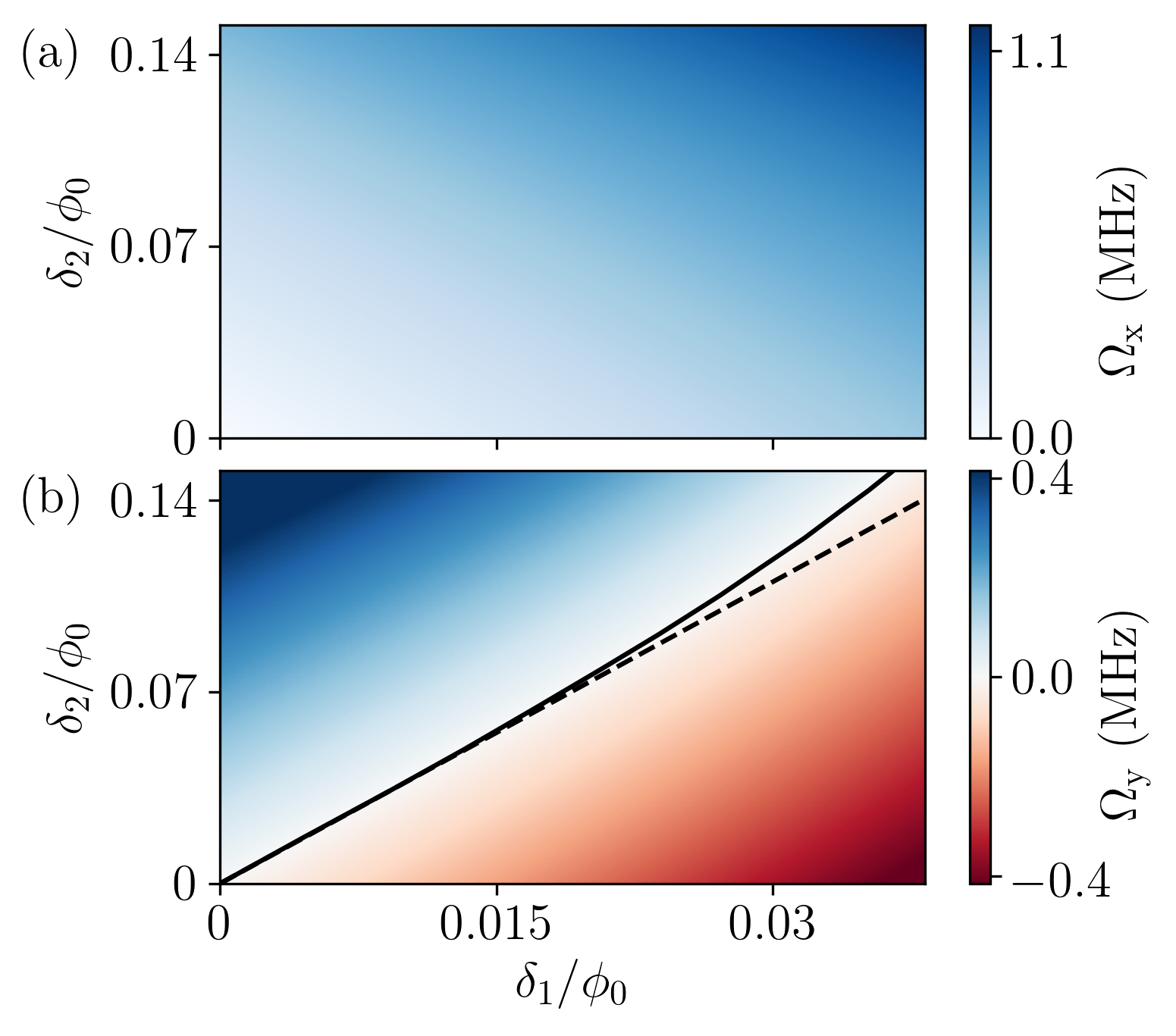}
	\caption{
		The coupling strengths $\Omega_{\rm x}$ (a) and $\Omega_{\rm y}$ (b) as a function of the modulation amplitudes $\delta_1$ and $\delta_2$. The values for $\Omega_{\rm x}$ and $\Omega_{\rm y}$ are obtained from  second order perturbation theory in $\delta_1$ and $\delta_2$ (the lengthy algebraic expression are not shown in the text). The dashed line is obtained from the first-order results in Eq.~(\ref{eq:omega_y_zero_condition}) and indicates where $\Omega_{\rm y}=0$. The solid line shows the corresponding second order results. 		
		The parameters are chosen as  $\omega_1/(2\pi)=5.8\,\rm GHz$, $\omega_2/(2\pi)=5.0\,\rm GHz$, $\omega_{\rm c}^0/(2\pi)=7.3\,\rm GHz$, $g_1/(2\pi)=g_2/(2\pi)=130\,\rm MHz$, $\theta=-0.1\,\phi_0$, $u_1/(2\pi)=u_2/(2\pi)=u_c/(2\pi)=-250\, \rm MHz$.}
	\label{fig:figure2}
\end{figure}

Driving the qubits on resonance and choosing the modulation amplitudes $\delta_1$ and $\delta_2$ such that $\Omega_{\rm y} = 0$ generates an effective Hamiltonian with pure Ising-type interaction ($\sim XX$). The phase $\varphi_j$ of the qubit drive determines the polarization of the transverse field. Choosing $\varphi_j = 0$ yields single qubit $\smx_j$ terms that correspond to the classical Ising model, whereas the choice $\varphi_j=-\pi/2$ leads to single qubit $\smy_j$ terms. Rotating about the x-axis by an angle of $\pi/2$ such that $\smy_j\rightarrow\smz_j$ then yields the transverse quantum Ising model
\begin{equation}
H_{\rm Ising} = \sum_j \frac{f_j}{2} \smz_j + \frac{\Omega_{\rm x}}{2}\smx_1\smx_2\,.\label{eq:classical_ising}
\end{equation}
Using Eq.~(\ref{eq:eff_iswap}) and Eq.~(\ref{eq:eff_bswap}), we can explicitly write the condition $\Omega_{\rm y}=0$ as
\begin{equation}
\frac{\delta_1}{\delta_2} = -\frac{\Delta_{1,-}^\theta\Delta_{2,+}^\theta+\Delta_{1,+}^\theta\Delta_{2,-}^\theta}{\Delta_{1,-}^\theta\Delta_{2,-}^\theta+\Delta_{1,+}^\theta\Delta_{2,+}^\theta}\,.\label{eq:omega_y_zero_condition}
\end{equation}
The dashed black line in Fig.~\ref{fig:figure2} (b) corresponds to the simple formula in Eq.~(\ref{eq:omega_y_zero_condition}) and agrees well with the results obtained from higher order pertubation theory for small modulation amplitudes.\\
\begin{figure}[tb]
	\centering
	\includegraphics[width=0.3\paperwidth]{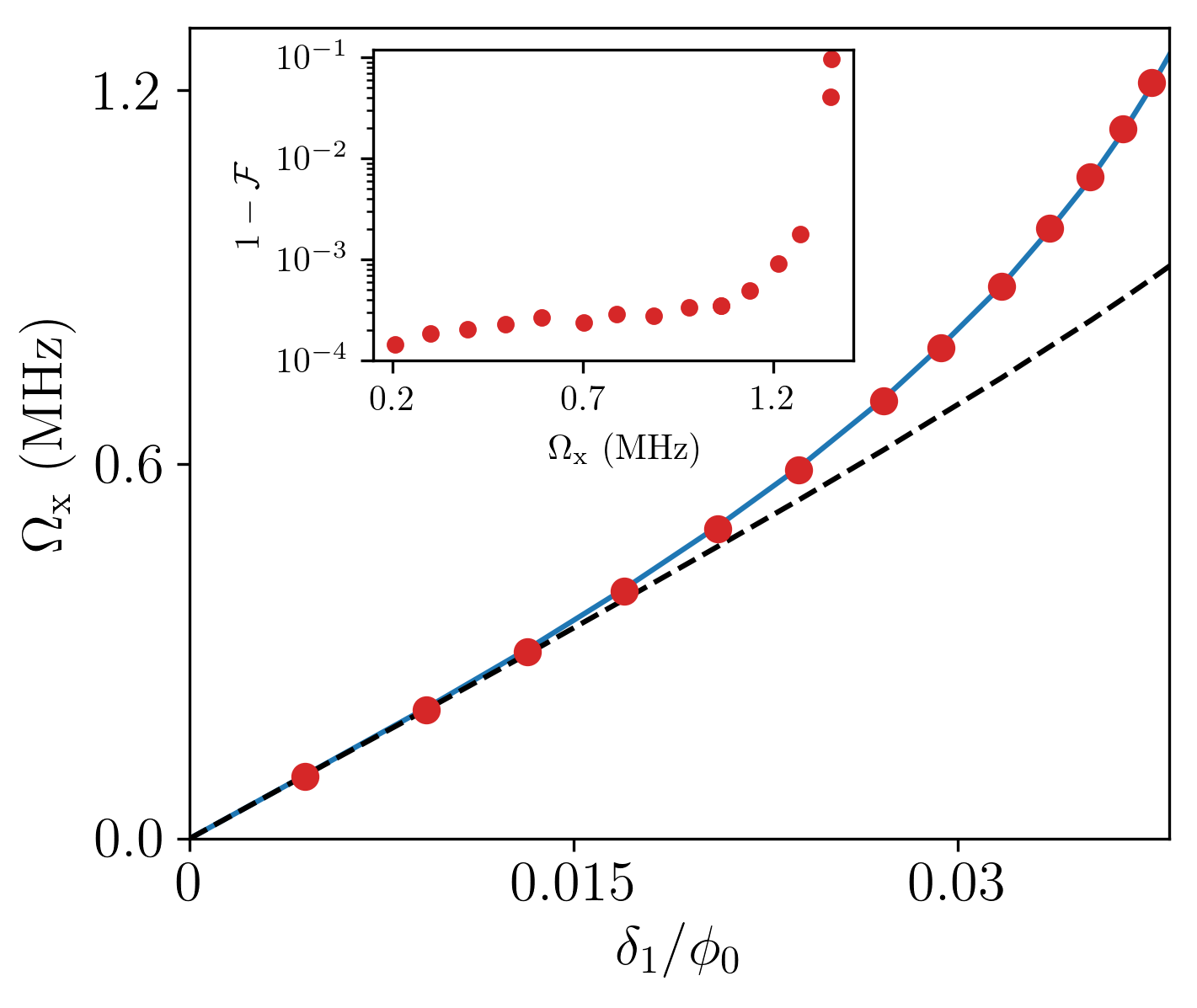}
	\caption{The coupling strength $\Omega_{\rm x}$ as a function of the modulation strength $\delta_1$. The value for $\delta_2$ has been chosen such that $\Omega_{\rm y}\approx 0$ for all values of $\delta_1$ (pure Ising interactions). The results of exact numerical simulations of the full transmon Hamiltonian~(\ref{eq:full_transmon_H}) (dots) are compared to analytical results obtained from perturbation theory in $\delta_1$ and $\delta_2$ to first [dashed line, see also Eqs.~(\ref{eq:eff_iswap}) and (\ref{eq:eff_bswap})] and second order (blue, solid). The inset shows the minimum infidelity $1-\mathcal{F}$ [Eq.~~(\ref{eq:average_fidelity})] for varying coupling strenghts. The parameters are chosen as in Fig.~\ref{fig:figure2}.}
	\label{fig:figure3}
\end{figure}
To benchmark our analytical findings for the coupling strength of the $XX$-interaction, we perform numerical simulations of the full transmon Hamiltonian  Eq.~(\ref{eq:full_transmon_H}) with the QuTiP package for Python \cite{JOHANSSON20131234}. In these simulations, the Hilbert space is truncated to three levels per transmon. The system is initialized in the four computational basis states and then time-evolved with the tunable coupler being modulated with different modulation amplitudes $\delta_1$ and $\delta_2$. In order to generated pure $XX$ interactions, we fix $\delta_1$ and choose $\delta_2$ according to the condition $\Omega_{\rm y} = 0$. The frequency of the resulting oscillations, i.e., the coupling strength, is then extracted numerically. As we are interested in the performance of the $XX$-gate $U_{xx} = e^{i\xi\smx_1\smx_2/2}$ for $\xi = \pi$, the qubit drives are turned off during the simulations, i.e., $f_1=f_2=0$. The results are shown in Fig.~\ref{fig:figure3}. A comparison with analytic results obtained from a perturbative expansion of the coupling strength in Eq.~(\ref{eq:effective_2qubithamiltonian}) in powers of the modulation amplitudes shows excellent agreement (solid and dashed line). The inset shows $1-\mathcal{F}$ with $\mathcal{F}=\max_t{\mathcal{F}(t)}$ and the average fidelity
\begin{equation}
\mathcal{F}(t)=\frac{1}{4}\sum_{\psi_0}\abs{\bra{\psi_0}U_{\rm xx}U_{\rm nsim}(t)\ket{\psi_0}}^2\,.
\label{eq:average_fidelity}
\end{equation}
Here, the sum runs over all computational basis states and $U_{\rm nsim}(t)\ket{\psi_0}\equiv\ket{\psi(t)}$ is the numerically time-evolved wave function. We conclude that for the chosen parameters, a coupling strength of up to $\sim 1.2$ MHz with an infidelity $1-\mathcal{F}$ below $\sim10^{-3}$ is readily achievable. For a stronger modulation, the fidelity of the gate decreases rapidly due to side-band excitations of the coupler \cite{Roth2017}. We also note that the sign of the Ising interaction strength can be chosen arbitrarily by adjusting the phase of the flux modulation in Eq.~(\ref{eq:phi_bi}).

\section{Adiabatic simulation in the rotating frame}
\label{sec:adiabatic_annealing}

\begin{figure}[tb]
	\centering
	\includegraphics[width=0.4\paperwidth]{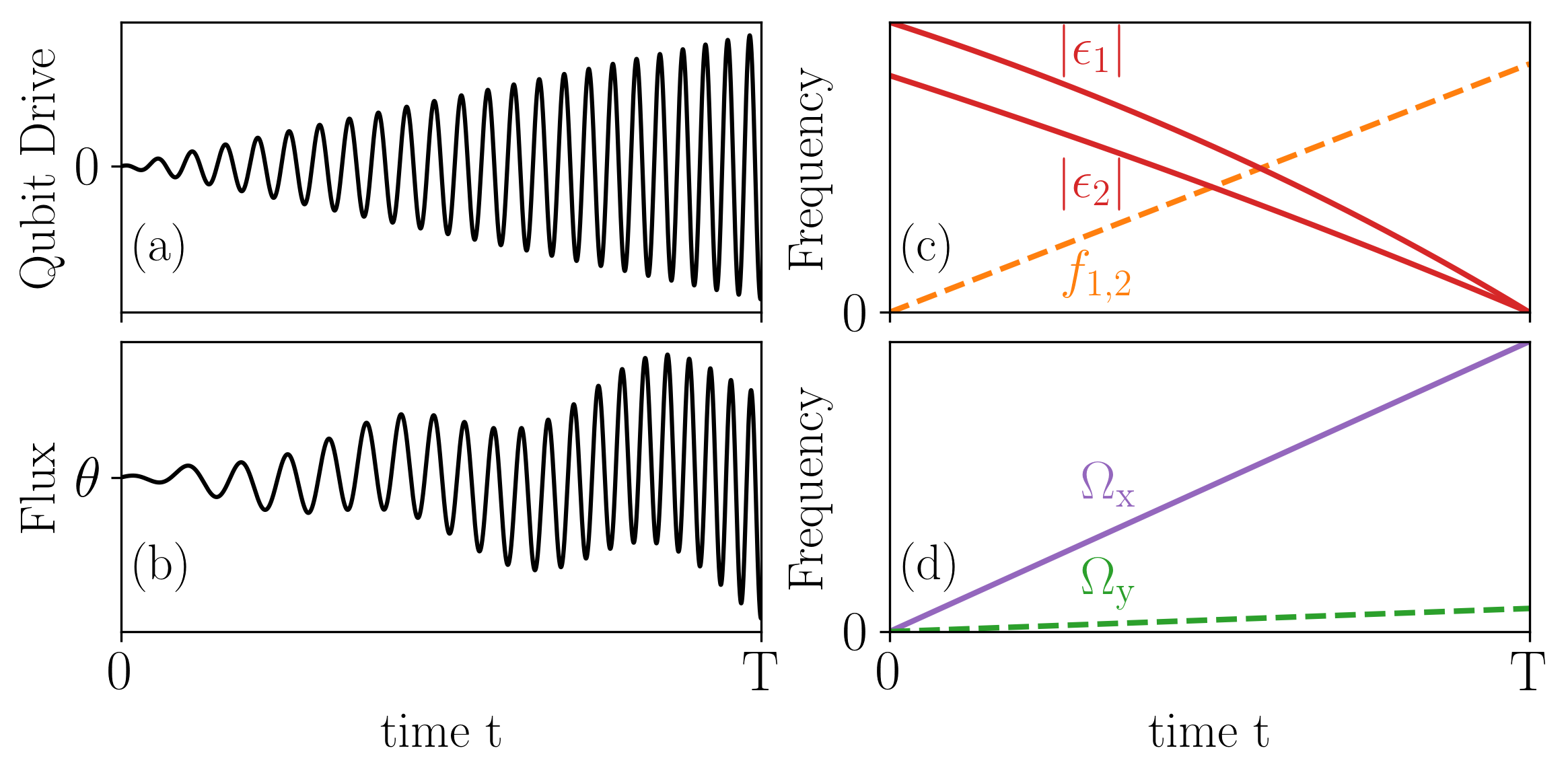}
	\caption{Schematic of a typical linear protocol as a function of time $t$ for an adiabatic simulation with single qubit terms and $XX$ and $YY$-interactions. The system is initialized in the ground state of Eq.~(\ref{eq:full_transmon_H}) with all external drives turned off, i.e., $f_j(0)=0$ and $\Phi(0)=\theta$. The qubits are subject to a harmonic microwave drive (a) with both, drive frequency $\od_j(t)$ [Eq.~(\ref{eq:adiabatic_drive_frequency})] and amplitude $f_j(t)$ [Eq.~(\ref{eq:adiabatic_f})] being varied linearly. (b) The magnetic flux amplitude threading the coupler SQUID loop is increased linearly with a bichromatic modulation with frequencies $\ophi_1(t)=\od_1(t)-\od_2(t)$ and $\ophi_2(t)=\od_1(t)+\od_2(t)$. Panels (c) and (d) show effective parameters of the Hamiltonian~(\ref{eq:effective_2qubithamiltonian}) for the simulation of the hydrogen molecule Eq.~(\ref{eq:h_h2}) for an atomic separation of $R= 0.37\,$\AA. The strength $f_j(t)$ (orange, dashed) of the transverse fields increases linearly. The shape of the detunings $\epsilon_{j}(t)$ (red, solid) is the result of the linear frequency variation of $\od_j(t)$ and an additional non-linear dispersive shift proportional to the modulation amplitudes $\delta_1(t)$ and $\delta_2(t)$. The dispersive shifts are obtained from second order pertubation theory (see Appendix~\ref{appendix_swt}). (d) Interaction strength $\Omega_{\rm x}$ (purple, solid) and $\Omega_{\rm y}$ (green, dashed) of the $XX$ and $YY$-interaction terms due to a linear increase of the flux modulation amplitude [see Eq.~\ref{adiabatic_modstrength}].}
	\label{fig:figure4}
\end{figure}

The parametric setup introduced in Section \ref{sec:model} can be used as an adiabatic quantum simulator. Here, the system is initialized in an eigenstate of an easily preparable Hamiltonian $H_0$ followed by an adiabatic variation of the Hamiltonian parameters. The final parameter configuration after a protocol run time $T$ is such that the system Hamiltonian matches some target Hamiltonian $H_{\rm T}$
\begin{equation}
H(t) = (1 - t/T) H_0 + \left(t/T\right) H_{\rm T} \, .
\end{equation}
We propose a scheme for the adiabatic simulation of Hamiltonian~(\ref{eq:generic_H}). It consists of three basic steps: initialization, time evolution and measurement. As in Sections~\ref{sec:effective_hamiltonian} and~\ref{sec:ising_interactions}, we focus on the special case of two qubits $N = 2$. A schematic of the adiabatic protocol can be found in Fig.~\ref{fig:figure4}.

First, as an initial Hamiltonian $H_0$, we choose the unmodulated and undriven Hamiltonian~(\ref{eq:full_transmon_H}) with $\omega_{\rm c} =\omega_{\rm c}^\theta$ and $f_1 = f_2 = 0$. Subsequently, the external drives and the modulation of the coupler are adiabatically turned on such that after a protocol runtime $T$, the effective Hamiltonian~\eqref{eq:effective_2qubithamiltonian} is equivalent to $H_{\rm T}$. For the qubit-qubit interactions, this is achieved by choosing the final values $\delta_1^{\rm T}$ and $\delta_2^{\rm T}$ such that they correspond to the interaction strengths $J_{1,2}^{\rm x}$ and $J_{1,2}^{\rm y}$ of $H_{\rm T}$ in Eq.~(\ref{eq:generic_H}). Using the analytical results in Eq.~(\ref{eq:eff_iswap}) and Eq.~(\ref{eq:eff_bswap}) this gives the conditions
\begin{equation}
\delta_1^{\rm T} = \frac{J_{1,2}^{\rm x}+J_{1,2}^{\rm y}}{g_1g_2\partial\omega_{\rm c}|_{\Phi=\theta}}\left(\frac{\Delta_{1,-}\Delta_{1,+}\Delta_{2,-}\Delta_{2,+}}{\Delta_{1,-}\Delta_{2,-}+\Delta_{1,+}\Delta_{2,+}}\right)\,,
\end{equation}
and
\begin{equation}
\delta_2^{\rm T} = \frac{J_{1,2}^{\rm x}-J_{1,2}^{\rm y}}{g_1g_2\partial\omega_{\rm c}|_{\Phi=\theta}}\left(\frac{\Delta_{1,-}\Delta_{1,+}\Delta_{2,-}\Delta_{2,+}}{\Delta_{1,-}\Delta_{2,+}+\Delta_{1,+}\Delta_{2,-}}\right)\,.
\end{equation}
The local bias fields can be simulated by generalizing the drive term used in Eq.~(\ref{eq:effective_hamiltonian_rot}) to incorporate explicitely time dependent frequencies
\begin{equation}
H_{\rm d}(t)=f_j(t)/2\left(\smp_j\exp(-i\left[\vartheta_j(t)+\varphi_j\right])+\hc\right)\,.
\end{equation}
Here, the drive strength $f_j(t)$ is potentially time-dependent as well. The transformation to the rotating frame is then mediated by 
\begin{equation}
R'(t)=\exp(-i\sum_{j=1,2}\vartheta_j(t)\smz_j/2)\,.
\end{equation}
As a result, the detuning $\epsilon_j$ in Eq.~\eqref{eq:effective_hamiltonian_rot} and Eq.~(\ref{eq:effective_2qubithamiltonian}) is replaced by the generalized expression 
\begin{equation}
\epsilon_j(t)=\od_j(t)-\bar{\omega}_i(t)\,,\label{eq:generalized_detuning}
\end{equation} 
where $\od_j(t)=\partial_t\vartheta_j(t)$. Here, $\bar{\omega}_i(t)$ denotes the effective qubit frequency, which is time-dependent because the dispersive shift depends on the modulation strength. An adiabatic variation of the $\smz_j$ terms can be thus accomplished by appropriatly varying the frequencies of the single qubit drives. The strength and polarization of the transverse fields can be controlled by the amplitude and phase of the qubit drives.

Finally, after this adiabatic variation step, the drives are turned off quickly and the final state of the system is measured. 
\subsection{Example: hydrogen molecule}
\begin{figure}[tb]
	\centering
	\subfloat[]{\includegraphics[width = 0.35\paperwidth]{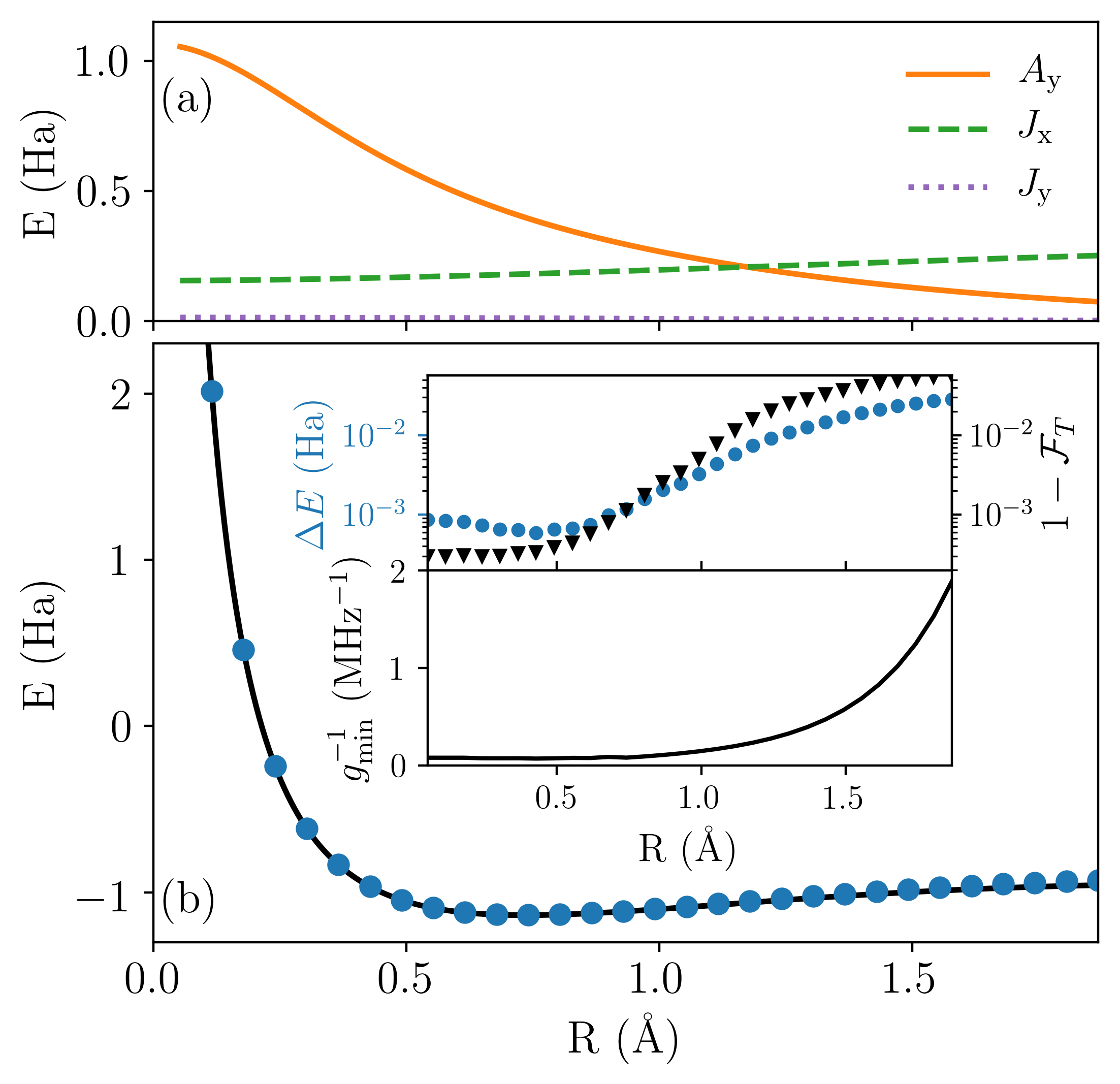}}
	
	\caption{ (a) Parameters of the Hamiltonian~(\ref{eq:h_h2}) of the hydrogen molecule as a function of the distance $R$ between the hydrogen atoms. The curves correspond to the final values in the adiabatic protocol and depict $A_{\rm y}$ (orange, solid), $J_{\rm x}$ (dashed, green) and $J_{\rm y}$ (dotted, purple). (b) Ground state energy of the hydrogen molecule as a function of $R$. The solid line  is obtained from numerical diagonalization of Eq.~(\ref{eq:h_h2}) whereas the dots (blue) are the results of numerical simulations with a fixed evolution time of $T=3.5\,\mu s$. The upper inset shows the difference $\Delta E$ between the energy obtained from numerical diagonalization of Eq.~(\ref{eq:h_h2}) and the simulation results (blue, dots), and the error $1-\mathcal{F}_{\rm T}$ of the fidelity at the end of the evolution (black, triangles). The lower inset shows the inverse of the minimal spectral gap during the adiabatic protocol. Here, we have chosen $\epsilon_0/(2\pi) = 2.5\,\rm MHz$ for all $R$ with the device parameters for the simulation being the same as in Fig.~\ref{fig:figure2}.} 
	\label{fig:figure5}
\end{figure}

A particular interesting application of adiabatic computing concerns the simulation of ground states of molecules in quantum chemistry. In the following, we apply the scheme described above to the simulation of the ground state of the hydrogen molecule and perform numerical simulations using realistic device parameters. The two-qubit target Hamiltonian
\begin{equation}
H_{\rm H_2}(R) = A_{\rm y}(R)\left(\smy_1 + \smy_2\right) + J_{\rm x}(R)\smx_1\smx_2 + J_{\rm y}(R)\smy_1\smy_2\,,\label{eq:h_h2}
\end{equation}
is isospectral to the hydrogen molecule in the singlet subspace \cite{Moll2016}. We effectively obtain this Hamiltonian by choosing $\varphi_j = -\pi/2$ as well as a vanishing detuning $\epsilon_j=0$ in Eq.~(\ref{eq:effective_2qubithamiltonian}). The precise parameter values for the drive and interaction strength then depend on the distance $R$ between the hydrogen atoms. Their spatial dependence is shown in Fig.~\ref{fig:figure5} (a). 
For small atomic distances, the eigenstates of this Hamiltonian are mostly governed by the single qubit terms, whereas with increasing atomic distances, the $XX$ interactions become dominant. 

We perform numerical simulations of the Schr\"odinger equation for the full parametric setup with Hamiltonian~(\ref{eq:full_transmon_H}) truncated to three levels for each transmon using QuTiP. An exemplary time evolution of the effective Hamiltonian parameters involved in the simulation of Eq.~(\ref{eq:h_h2}) is shown in Fig.~\ref{fig:figure4} (c) and (d). For simplicity, we choose a linear adiabatic protocol for all parameters. The drive and modulation strengths are tuned according to
\begin{eqnarray}
f_1(t)=f_2(t) &=& A_{\rm y}(R)\frac{t}{T}\,,\label{eq:adiabatic_f}
\end{eqnarray}
and
\begin{eqnarray}
\label{adiabatic_modstrength}
\delta_m(t) &=&\delta_m^{\rm T}(R)\frac{t}{T}\,,
\end{eqnarray}
where $\delta^{\rm T}_m(R)$ is chosen such that $\Omega_{\rm x}(T) = J_{\rm x}(R)$ and $\Omega_{\rm y}(T) = J_{\rm y}(R)$.
The protocol for the drive frequency $\od_j(t)$ is given by
\begin{align}
\od_j(t)= \bar{\omega}_j^T-\epsilon_0\left(1-\frac{t}{T}\right)\,.\label{eq:adiabatic_drive_frequency}
\end{align}
The second term in Eq.~\eqref{eq:adiabatic_drive_frequency} ensures a small initial qubit detuning at $t=0$ in order to mimic a $Z$-polarization of the initial Hamiltonian in the rotating frame. The effective qubit frequencies $\bar{\omega}_j^{\rm T} = \bar{\omega}_i(T)$ at the end of the evolution are obtained by numerically maximizing the contrast of the oscillations between the $\ket{01}$ and $\ket{10}$, and the $\ket{00}$ and $\ket{11}$ states while simultaneously modulating the coupler at both final values $\delta_{1}^{\rm T}$ and $\delta_{2}^{\rm T}$ of the modulation amplitudes. The effective qubit frequencies are then given by half of the sum/difference of the respective transition frequencies. 
In order to ensure that the interaction terms in Eq.~(\ref{eq:effective_hamiltonian_rot}) are static in the rotating frame, the coupler modulation frequencies are kept in resonance with the sum and difference of the qubit drive frequencies throughout the evolution. In Eq.~(\ref{eq:adiabatic_drive_frequency}) we have chosen $\epsilon_0$ to be the same for both qubits such that the difference modulation frequency $\ophi_1(t) = \bar{\omega}_1^{\rm T} - \bar{\omega}_2^{T}$ is constant during the protocol whereas the sum modulation frequency $\ophi_2(t)$ is given by $\ophi_2(t) = \bar{\omega}_1^{\rm T} + \bar{\omega}_2^{\rm T} - 2\epsilon_0 \left(1 - t/T\right)$. In the numerical simulations, we have fixed $\epsilon_0/(2\pi)=2.5\,\rm MHz$.

At the end of the evolution, we extract a two-qubit state $\psi_{\rm T}$ from the three-transmon state $\Psi_{\rm T}$ by performing a partial trace over the coupler degrees of freedom followed by a projection to the two lowest qubit energy levels. The resulting state is then transformed to the rotating frame $\widetilde{\psi}_{\rm T}=R'(T)\psi_{\rm T}R'^\dag(T)$. We then calculate the energy for each value of $R$ by determining the expectation values $\expval{\smy_1}$, $\expval{\smy_2}$, $\expval{\smy_1\smy_2}$ and $\expval{\smx_1\smx_2}$. Here, the expectation values are taken with respect to the final state $\widetilde{\psi}_{\rm T}$. The energy is given by
\begin{equation}
E=\expval{H_{\rm T}}=A_{\rm y}\left(\expval{\smy_1} + \expval{\smy_2}\right) + J_{\rm x}\expval{\smx_1\smx_2} + J_{\rm y}\expval{\smy_1\smy_2}\, . \label{eq:energy}
\end{equation}

Fig.~\ref{fig:figure5} (b) shows the ground state energy of the hydrogen molecule. For the results of the numerical simulations (dots), the protocol time is kept fixed at $T = 3.5\,\mu s$ for all values of $R$. The respective final values for the modulation amplitudes $\delta_1^{\rm T}$ and $\delta_2^{\rm T}$ are obtained from second order pertubation theory in $\delta_1$ and $\delta_2$. The minimal evolution time required for an adiabatic evolution is inversely proportional to the minimal gap $g_{\rm min}$ between the ground state and the first excited state. The inverse $g_{\rm min}^{-1}$ of this spectral gap for the effective Hamiltonian Eq.~(\ref{eq:effective_2qubithamiltonian}) can be seen in the lower inset of Fig.~\ref{fig:figure5} (b). We find that the simulation results are in very good agreement with the exact ground state for atomic distances below $\sim 0.8\,\rm\AA$. The difference $\Delta E$ between the energy obtained from the simulations and the numerical diagonalization of Eq.~(\ref{eq:h_h2}) is well below chemical accuracy ($< 1.5\, \rm mHa$). In this regime, the final Hamiltonian is dominated by single qubit terms and the corresponding spectral energy gap is relatively large. Finally, we calculate the final fidelity 
\begin{equation}
\mathcal{F}_{\rm T} = \Tr\left(\sqrt{\sqrt{\widetilde{\psi}_{\rm T}}\phi_{\rm T}\sqrt{\widetilde{\psi}_{\rm T}}}\right)^2\,,
\end{equation}
with $\phi_{\rm T}= \dyad{\phi_T}$ being the ground state of the target Hamiltonian obtained from numerical diagonalization of Eq.~(\ref{eq:h_h2}). Both, $\Delta E$ and the error $1-\mathcal{F}_{\rm T}$ increase notably for an atomic distance larger than $\sim 0.8\,\rm\AA$. This is due to the decreasing spectral gap with increasing $R$. Achieving the same level of precision as for smaller values of $R$ requires slower sweep speeds which correspond to longer evolution times for a fixed protocol.
\subsection{Dissipative protocol}
In order to take dissipative effects into account, we simulate the Lindblad master equation
\begin{equation}
\label{eq:master}
\dot\Psi=-i[H_{\rm Tr},\Psi]+\sum_{j = 1,2, \rm c} \left(\Gamma_{j}^-\mathcal{L}[a_j]\Psi + \Gamma_j^{\rm z}\mathcal{L}[a^\dag_j a_j]\Psi\right),
\end{equation}
with the standard Lindblad operator 
\begin{equation}
\mathcal{L}[C] = (2 C \Psi(t) C^{\dag} -\acomm{\Psi(t)} {C^{\dag}C})/2\,.
\end{equation}
Here, $\acomm{\cdot}{\cdot}$ denotes the anticommutator. The decoherence rates are determined in terms of coherence times $T_{1,j}$ and $T_{2,j}$ ($j= 1, 2, \rm c$) via $\Gamma_j^{\rm z} = (1/2)(1/T_{2,j} - 1/(2T_{1,j}))$ and $\Gamma_{j}^- = 1/T_{1,j}$. 

\begin{figure}[tb]
	\centering
	\subfloat[]{\includegraphics[width = 0.4\paperwidth]{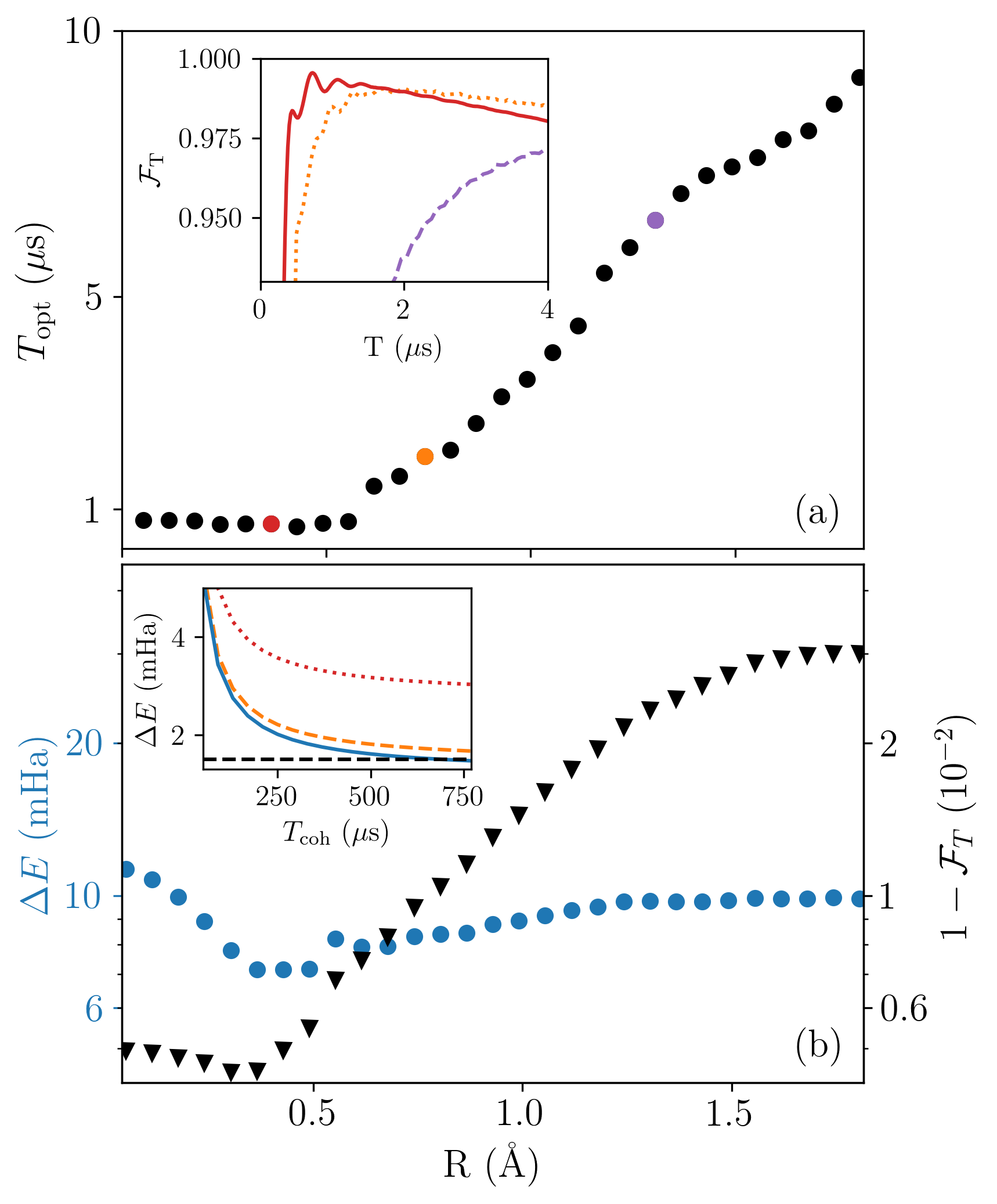}}

	\caption{ (a) Optimal run time $T_{\rm opt}$ of the adiabatic protocol for the ground state of the hydrogen molecule as a function of atomic distance $R$. Results are numerical solutions of the master equation~(\ref{eq:master}). The inset shows the final fidelity $\mathcal{F}_{\rm T}$ as a function of the protocol run time $T$ for $R = 0.37\,\rm\AA$ (red, solid), $R = 0.74\,\rm\AA$ (orange, dotted) and $R = 1.30\,\rm\AA$ (purple, dashed). The corresponding data points in the main plot are highlighted with the respective colors. (b) Error $\Delta E$ of the energy (blue) and error $1-\mathcal{F}_{\rm T}$ of the final fidelity (black)  at the optimal protocol runtimes $T_{\rm opt}$ for varying values of $R$. The Hamiltonian parameters are chosen as in Fig.~\ref{fig:figure2} with an initial detuning of $\epsilon_0/(2\pi) =2.5\,\rm MHz$. Decoherence times are $T_{1,1}=T_{1,2} = 60\,\mu s$, $T_{2,1}=T_{2,2} = 40\,\mu s$,  $T_{1,\rm c}= 10\,\mu s$ and $T_{2,\rm c} = 1\,\mu s$ of qubit $1$, qubit $2$ and the tunable coupler, respectively. The inset in (b) shows $\Delta E$ for $R=0.37\,\text{\AA}$ for a fixed protocol run time $T = 730\,\rm ns$ as function of coherence times $T_{\rm coh} \equiv T_{1,j}=T_{2,j}$ for ($j=1,2$) of the qubits. The solid line (blue) shows the result for a decoherence-free coupler $\Gamma_{\rm c}^z =\Gamma_{\rm c}^-=0$. The dashed (orange) and dotted (red) lines are both obtained with the same $T_{1,\rm c}$ and different $T_{2,\rm c}$, i.e., $T_{1,\rm c}=10\,\mu s$ and $T_{2,\rm c}=10\,\mu s$ and $T_{2,\rm c}=1\,\mu s$, respectively. The chemical accuracy threshold is shown as a horizontal line (black, dashed).}
	\label{fig:figure6}
\end{figure}

 Fig.~\ref{fig:figure6} shows results of simulations performed with the same device parameters as in the caption of Fig.~\ref{fig:figure2}. The dissipation rates $\Gamma_j^-$ and $\Gamma_j^{\rm z}$, stated in the caption of Fig.~\ref{fig:figure6}, are chosen to have realistic values for currently used devices.
In contrast to the dissipation-free simulations, there are two competing processes that lead to the emergence of an optimal protocol run time $T_{\rm opt}$ at which $\mathcal{F}_{\rm T}$ is maximal and $\Delta E$ is minimal. 
Short protocol times, i.e., fast sweep rates lead to a decrease in the fidelity because of non-adiabatic transitions whereas for long protocol times, i.e., slow sweep rates, a decrease in fidelity is caused by dissipation and decoherence. This determines the optimal protocol run times $T_{\rm opt}$.
 Note that in Eq.~(\ref{eq:master}) we implicitly assume a zero-temperature bath. In non-driven systems, this leads to a relaxation into the instantaneous ground state \cite{Albash2015}. This is in contrast to the case considered here where the external driving leads to a non-equlibrium steady state.\\
  The optimal protocol time as a function of atomic separation as obtained by numerical simulation of the adiabatic protocol for values of $T$ ranging from $0.4\,\mu s$ to $10\,\mu s$ is shown in Fig.~\ref{fig:figure6} (a). The optimal time $T_{\rm opt}$ remains approximately constant at about $\sim 1\,\rm\mu s$ for atomic distances below $R \sim 0.55\,\rm\AA$ and increases as $R$ is increased further. This is in agreement with the findings for the dissipation-less protocol. The decreasing gap for large values of $R$ requires slow adiabatic sweep speeds to reduce the effects of non-adiabatic transitions. As can be seen in the inset of Fig.~\ref{fig:figure6} (a), the fidelity as a function $T$ features oscillation for short evolution times in the regime of small $R$ which are smoothened for larger values of $R$. The shape of these oscillations is strongly dependent on the protocol \cite{PhysRevA.87.012116} and the precise location of their maxima determines the exact values of $\Delta E$ and $\mathcal{F}_{\rm T}$. As can be seen in Fig.~\ref{fig:figure6} (b), the fidelity and the accuracy in determining $E$ decrease with increasing $R$ similar to the dissipation-free case. Notably however, they feature a maximum around $\sim 0.3\,\rm\AA$ (corresponding to a minimum in $\Delta E$ and $1 - \mathcal{F}$).\\ To study the influence of the different decoherence channels in more detail, we perform simulations with varying qubit and coupler coherence times. The inset in Fig.~\ref{fig:figure6} (b) shows the error $\Delta E$ of the energy as a function of coherence times $T_{\rm coh}$ of the qubits where for simplicity, the $T_1$ and $T_2$ times are chosen to be the same, i.e., $T_{\rm coh} \equiv T_{1,j}=T_{2,j}$ for ($j=1,2$). In this example, the atomic separation is $R=0.37\,\text{\AA}$ with a protocol run time of $730\,\rm ns$. We compared the results obtained with an ideal coupler with $\Gamma_{\rm c}^z = \Gamma_{\rm c}^- = 0$ to simulations with a coupler coherence time of $T_{1,\rm c}= 10\,\mu s$ and two different $T_2$ times $T_{2,\rm c}=1\,\mu s$ (red, dotted) and $T_{2,\rm c}=10\,\mu s$ (orange, dashed). We find that chemical accuracy can be reached at a qubit coherence time of $T_{\rm coh}\sim 650\,\mu s$ for the dissipation-free coupler. With the same qubit coherence times, the results obtained from a dissipative coupler with $T_{2,\rm c} = 1\,\mu s$ are $\sim 2\,\rm mHa$ away from chemical accuracy, whereas with a coupler coherence time of $T_{2,\rm c} =10\,\mu s$ the error $\Delta E$ is only $\sim 0.2\,\rm mHa$ away from chemical accuracy. We therefore conclude that the dominant coherence time scale in the adiabatic protocol is the coherence time of the computational qubits  if the coupler coherence is longer than the protocol duration $T$. This is in agreement with the results obtained for the gate fidelity of iSWAP and bSWAP gates in Ref.~\cite{Roth2017}.
 

%
\section{Summary and outlook}
\label{sec:conclusion}

In summary, we have theoretically studied a parametric modulation scheme based on the polychromatic modulation of a tunable coupling device that allows for the creation of pure Ising-type interactions as well as a combination of $XX$ and $YY$-type interactions with an arbitrary ratio. We have derived compact analytic expression for the resulting coupling strength that are in good agreement with numerical calculations. In addition, it is possible to obtain a $ZZ$-term by driving the $\ket{11}\leftrightarrow\ket{20}$ transition off-resonantly \cite{Kounalakis2018, Reagor2018}, which in combination with the $XX$ and $YY$-terms can be used to implement more general Hamiltonians as subject of further studies.\\
By considering the hydrogen molecule as an example system, we have numerically demonstrated the feasibility of performing adiabatic quantum simulations with parametrically generated interactions. The proposed simulations are performed in a rotating frame with an external drive applied to the qubits. For small atomic distances, we are able to calculate the molecular energy with chemical accuracy in dissipation-free simulations. The predicted optimal protocol run time for dissipative simulations is in the few microsecond range for typical device and coherence parameters. We find that for the linear protocol used in this work, coherence times of a few hundred microseconds are needed to reach chemical accuracy. We suggest that using an optimized, rather than a linear adiabatic protocol, is a suitable measure to reduce the necessary protocol run time in order to reduce dissipative effects. The presented scheme is generalizable to the simulation of excited states by choosing different initial states. Furthermore, the model is also valid for more than two qubits coupled to a single tunable coupler. In a future step, protocols for multi-qubit systems should be investigated. They would allow simulating more complex Hamiltonians. While for simple systems such as the hydrogen molecule the required interactions can be implemented directly, for more complicated molecules interactions may have to be decomposed into 2-local interactions using perturbative gadgets. However, our method is directly applicable to study interacting spin systems with nearest-neighbour couplings with adjustable ferromagnetic and anti-ferromagnetic couplings.
\begin{acknowledgments}
We thank David P. DiVincenzo for insightful disccusions. This work was supported by the IARPA LogiQ program under contract W911NF-16-1-0114-FE and the ARO under contract W911NF-14-1-0124.
\end{acknowledgments}
\appendix
\section{Time-dependent Schrieffer-Wolff transformation (SWT)}
\label{appendix_swt}

We briefly summarize the main results of the time-dependent SWT derived in \cite{Roth2017}. First, we separate the Hamiltonian~(\ref{eq:h_three_qubits}) into a local part $H_0$ and two-qubit interactions $V$, i.e., we write $H = H_0 + V$ with $V=\sum_j g_j\smx_j\smx_{\rm c}$. This Hamiltonian can be brought into block diagonal form by applying a unitary transformation $U(t)=\exp{S(t)}$ with $S^\dag=-S$. A perturbative expansion to second order in the coupling $g_j$ yields
\begin{equation}
\bar{H}=UHU^\dag-i U\left(\pdv{U}{t}\right)^\dag\approx H_{\rm eff} + H_{\rm V}\,,
\end{equation}
with
\begin{equation}
H_{\rm eff} \approx H_0 +\comm{S}{V} + \half\comm{S}{\comm{S}{H_0}}+\frac{i}{2}\comm{S}{\pdv{S}{t}}\,,\label{eq:h_eff_commutators}
\end{equation}
and
\begin{equation}
H_{\rm V}=i\pdv{S}{t}+\comm{S}{H_0}+V\,.
\end{equation}
Inserting the ansatz Eq.~(\ref{eq:ansatz_S}) into Eq.~(\ref{eq:h_eff_commutators}) yields
\begin{align}
\Heff&=\sum_{j}\left[ -\frac{\bar{\omega}_j}{2}\sigma^z_j +  \left(  \frac{f_j}{2}e^{-i \left( \od_j t + \varphi_j \right)} \smp_j  + {\rm h.c.} \right) \right] 
		 \nonumber\\
&- \frac{\bar{\omega}_{\rm c}(t)}{2}\smz_{\rm c} + \sum_{i<j} \Big(\Omega^+_{ij}(t)\smp_i\smm_j \smz_{\rm c}\nonumber\\
 &+ \Omega^-_{ij}(t)\smp_i\smp_j \smz_{\rm c} +\hc\Big)\,.
\label{eq:heff_dispersive}
\end{align}
Here, the dispersively shifted qubit frequencies 
\begin{align}
\bar{\omega}_j(t)& = \omega_j + g_j \sum_{\mu=\pm}\Re(\alpha_{j,\mu}(t))\,, 
\label{eq:dispersive_shifts_qubits}
\end{align}
and the coupling strengths
\begin{align}
\label{eq:J_delta}
\Omega^-_{ij}(t)  =   \Big( &g_i \alpha_{j,-}^*(t) + g_j \alpha_{i,-}(t)   \nonumber\\
 -&g_i \alpha_{j,+}^*(t) - g_j \alpha_{i,+}(t)\Big)/2\,,
\end{align}
as well as
\begin{align}
\Omega^+_{ij}(t) =  \Big(&g_i \alpha_{j,-}(t) + g_j \alpha_{i,-}(t)  \nonumber\\
  -&g_i \alpha_{j,+}(t) - g_j \alpha_{i,+}(t) \Big)/2\,,
\label{eq:J_sigma}
\end{align}
are functions of the Schrieffer-Wolff coefficients $\alpha_{j,\pm}(t)$. The necessary condition that the qubit and coupler subspaces decouple is  $H_{\rm V} = 0$. This leads to the differential Eq.~(\ref{eq:deq_alpha}) from which the coefficients $\alpha_{j,\pm}(t)$ can be determined (see Appendix~\ref{app:alpha_fourier} for an explicit solution). 
In arriving at Eq.~(\ref{eq:heff_dispersive}) we assume that the drive is left invariant under the SWT (see Appendix~\ref{appendix_drive}).We now set $\smz_c\approx 1$ in Eq.~(\ref{eq:heff_dispersive}) and omit the resulting constant term. A subsequent transformation to a frame rotating at the qubit drive frequencies yields the effective two-qubit Hamiltonian~(\ref{eq:effective_hamiltonian_rot}) in the main text. 

\section{Discussion of the drive terms}
\label{appendix_drive}

The generator of the SWT in Eq.~(\ref{eq:ansatz_S}) is chosen such that the transformed Hamiltonian is block diagonal in the absence of external driving ($f_j=0$). In order to ensure that this ansatz is still applicable for small, but finite drive amplitudes, we now consider the dispersive transformation of the drive Hamiltonian $H_{\rm d}=\sum_j (f_j/2) \smp e^{-i\od_j t} + {\rm h.c.}$ (here, we have chosen $\varphi_1=\varphi_2=0$ for simplicity). To leading order in $g_j$ we need to consider
$\widetilde{H}^{\rm d} \approx H^{\rm d} + \comm{S}{H^{\rm d}} $. The commutator yields terms of the type $\sim f_j \alpha_{j,\pm}(t) \smz_j\sigma^\pm_{\rm c}$, which in principle lead to a coupling between the computational and the tunable coupler. 
However, as long as $f_j \ll g_j$ these terms are still small compared to the effective interactions $\Omega_{j,\pm} \sim g_j \alpha_{j,\pm}$.
With the external drives near resonance and modulation frequencies chosen as in the main text, 
the residual coupling terms still rotate rapidly at the side-band frequencies $\sim\abs{\Delta_{j,\pm}}$ and can thus be neglected.
In this paper we have chosen bare coupling strength $g_j$ which exceed the drive strength $f_j$ by more than two orders of magnitude 
and find excellent agreement between the effective model in Eq.~(\ref{eq:effective_2qubithamiltonian}) and the exact numerics based on Eq.~(\ref{eq:full_transmon_H}), see Fig.~\ref{fig:figure5}.

\section{SWT coefficients for bichromatic modulations}
\label{app:alpha_fourier}
Eq.~(\ref{eq:deq_alpha}) is of the general form
\begin{equation}
\dot{y}+P(t)y=Q\,,
\end{equation}
with $y=\alpha_{j,\pm}(t)$, $Q=ig_j$ and $P(t)=\Delta_{j,\pm}(t)$. It has the solution
\begin{equation}
y(t)=u^{-1}(t)\left(\int u(t')Q\dd{t'}+C_{\pm}\right)\,,
\end{equation}
with the integrating factor $u(t)=\exp(\int P(t')\dd{t'})$ and the integration constant $C_\pm$. For the bichromatic modulation Eq.~(\ref{eq:phi_bi}) we expand the coupler frequency to first order in $\delta_1$ and $\delta_2$ [Eq.~(\ref{eq:otbexpand_bichromatic})] which yields the integrating factor
\begin{equation}
u(t)=e^{i\Delta^\theta_{j,\pm}t}e^{i \pm\sum_m \lambda_m\sin(\ophi_mt)}\,,
\end{equation}
with $\lambda_m=\partial_\Phi\otb \big|_{\Phi=\theta}\delta_m/\ophi_m$ and the detuning $\Delta^\theta_{j,\pm}=\omega_j \pm \omega_{\rm c}^\theta$.
Using the identity
\begin{equation}
e^{ix\sin{z}}=\sum_{n=-\infty}^\infty J_n(x)e^{iz}\,,\label{eq:jacobi_anger}
\end{equation}
where $J_n(x)$ is the $n$-th Bessel function of the first kind, we can solve the remaining integral. For the case of a bichromatic modulation ($M=2$)
we find
\begin{align}
\alpha_{j,\pm}(t)&= g_j\sum_{m,n,n',m'=-\infty}^\infty J_{m-m'}\left(\mp\lambda_1\right)J_{n-n'}\left(\mp\lambda_2\right)\nonumber\\
\cross&J_{m'}\left(\pm\lambda_1\right)J_{n'}\left(\pm\lambda_2\right)
\frac{e^{i\left(m\ophi_1+n\ophi_2\right)t}}{m'\ophi_1+n'\ophi_2+\Delta^\theta_{i,\pm}}\nonumber\\
&+C_{j,\pm}(t)\,,\label{eq:alpha_complete}
\end{align}
where we have defined
\begin{equation}
C_{j,\pm}(t) = e^{-i\Delta^\theta_{j,\pm}t}e^{\pm i\sum_m \lambda_m\sin(\ophi_mt)}C^0_{j,\pm}\,.\label{eq:C(t)}
\end{equation}
With the initial condition $\alpha_{j,\pm}(0)=g_j/\Delta^\theta_{j,\pm}$, we obtain
\begin{align}
C^0_{j,\pm}& = \frac{g_j}{\Delta_{j,\pm}^\theta}-g_j\sum_{\substack{m,n,m',n'}}\frac{J_{m'}(\pm\lambda_1)J_{n'}(\pm\lambda_2)}{m'\ophi_1+n'\ophi_2+\Delta^\theta_{j,\pm}}\,.
\end{align}
As in the main text, we consider near resonant qubit driving $\od_j\approx\bar{\omega}_j$ and choose the modulation frequencies to correspond to the sum and the difference of the qubit drive frequencies, i.e., $\ophi_1\approx\od_1-\od_2$ and $\ophi_2\approx\od_1+\od_2$. In order to generate static interactions in Eq.~(\ref{eq:effective_2qubithamiltonian}), we only keep terms in Eq.~(\ref{eq:alpha_complete}) that rotate at multiples of $\ophi_1$ {\emph or} $\ophi_2$. The only terms satisfying this conditions are those where either $m=0, n \neq 0$ or $n=0, m\neq 0 $, i.e., 
\begin{equation}
\alpha_{j,\pm}(t) \approx  \bar{\alpha}_{j,\pm}(0) + \sum_{m,k\neq 0} \bar{\alpha}_{j,\pm}(k,m)e^{ik\ophi_mt} \,,\label{eq:alpha_appendix}
\end{equation}
with $\bar{\alpha}_{j,\pm}(k,m)$ given in Eq.~(\ref{eq:alpha_bar_appendix}) as
\begin{align}
\bar{\alpha}_{j,\pm}(k,m) &= g_j\sum_{q,p} 
\frac{ J_{k-q}\left(\mp\lambda_m\right)J_q\left(\pm\lambda_m\right) J_{p}\left(\pm\lambda_{n}\right)^2}{q\ophi_m+p\ophi_{n}+\Delta^\theta_{j,\pm}}\,,\label{eq:alpha_bar_appendix}
\end{align}
where $n\neq m$ and
	\begin{equation}
\bar{\alpha}_{j,\pm}(0) =g_j\sum_{q,p} 
\frac{J_q\left(\pm\lambda_1\right)^2 J_{p}\left(\pm\lambda_{2}\right)^2}{q\ophi_1+p\ophi_{2}+\Delta^\theta_{j,\pm}}\,.\label{eq:alpha_bar_zero}
\end{equation}
Eq~(\ref{eq:alpha_appendix}) is then plugged into the corresponding expressions Eq.~(\ref{eq:J_delta}) and Eq.~(\ref{eq:J_sigma}) for the coupling strength whereas substituting Eq.~(\ref{eq:alpha_appendix}) into Eq.~(\ref{eq:dispersive_shifts_qubits}) yields the dispersive shifts.\\
 By expanding the coupler frequency up to second order in the modulation strength $\delta_m$ one can still solve Eq.~(\ref{eq:deq_alpha}) and derive analytic expression for the coupling strength and dispersive shifts. Since these expressions are rather cumbersome to write down, we have omitted them here for brevity.

\section{Weak modulation approximation}
\label{app:weakmodulation}

 We can simplify Eq.~(\ref{eq:alpha_bar_appendix}) and Eq.~(\ref{eq:alpha_bar_zero}) in the limit of small modulation strength $\delta_m$ such that $\lambda_m\ll1$. In this limit, the Bessel functions can be approximated by $J_n(x) \approx (1/n!)(x/2)^n$ for $n\geq0$ and $J_{n}(x)\approx 1/(-n)!(-x/2)^{-n}$ for $n<0$. 
First, we consider the term with $k=0$ [Eq.~(\ref{eq:alpha_bar_zero})]. To leading order in $\lambda_m$ we obtain	
\begin{align}
&\bar{\alpha}_{j,\pm}^j(0) = \frac{g_j}{\Delta^\theta_{j,\pm}}+\sum_{m'=1,2}\frac{g_j}{2}\frac{\lambda_{m'}^2\Delta^\theta_{j,\pm}}{(\Delta^\theta_{j,\pm})^2-(\ophi_{m'})^2}+\mathcal{O}(\lambda^3_m)\,.
\end{align}
Plugging this result into Eq~(\ref{eq:alpha_appendix}) and subsequently into  Eq.~(\ref{eq:dispersive_shifts_qubits}) yields an analytic expression for the dispersive shifts.
Finally, we evaluate the term with $\abs{k}=1$ in Eq.~(\ref{eq:alpha_bar_appendix}) and obtain
\begin{align}
\bar{\alpha}_{j,\pm}(k,m)
=\pm \frac{g_j}{2}k\Big[\frac{\lambda_m}{k\ophi_m+\Delta^\theta_{j,\pm}}-\frac{\lambda_m}{\Delta^\theta_{j,\pm}}\Big]+\mathcal{O}(\lambda^3_m)\,.
\end{align}
Plugging this result into Eq.~(\ref{eq:Omega_bar}) yields the effective coupling strengths Eq.~(\ref{eq:eff_iswap}) and Eq.~(\ref{eq:eff_bswap}) in the main text.

\bibliography{literature}

\begin{thebibliography}{50}%
\makeatletter
\providecommand \@ifxundefined [1]{%
 \@ifx{#1\undefined}
}%
\providecommand \@ifnum [1]{%
 \ifnum #1\expandafter \@firstoftwo
 \else \expandafter \@secondoftwo
 \fi
}%
\providecommand \@ifx [1]{%
 \ifx #1\expandafter \@firstoftwo
 \else \expandafter \@secondoftwo
 \fi
}%
\providecommand \natexlab [1]{#1}%
\providecommand \enquote  [1]{``#1''}%
\providecommand \bibnamefont  [1]{#1}%
\providecommand \bibfnamefont [1]{#1}%
\providecommand \citenamefont [1]{#1}%
\providecommand \href@noop [0]{\@secondoftwo}%
\providecommand \href [0]{\begingroup \@sanitize@url \@href}%
\providecommand \@href[1]{\@@startlink{#1}\@@href}%
\providecommand \@@href[1]{\endgroup#1\@@endlink}%
\providecommand \@sanitize@url [0]{\catcode `\\12\catcode `\$12\catcode
  `\&12\catcode `\#12\catcode `\^12\catcode `\_12\catcode `\%12\relax}%
\providecommand \@@startlink[1]{}%
\providecommand \@@endlink[0]{}%
\providecommand \url  [0]{\begingroup\@sanitize@url \@url }%
\providecommand \@url [1]{\endgroup\@href {#1}{\urlprefix }}%
\providecommand \urlprefix  [0]{URL }%
\providecommand \Eprint [0]{\href }%
\providecommand \doibase [0]{http://dx.doi.org/}%
\providecommand \selectlanguage [0]{\@gobble}%
\providecommand \bibinfo  [0]{\@secondoftwo}%
\providecommand \bibfield  [0]{\@secondoftwo}%
\providecommand \translation [1]{[#1]}%
\providecommand \BibitemOpen [0]{}%
\providecommand \bibitemStop [0]{}%
\providecommand \bibitemNoStop [0]{.\EOS\space}%
\providecommand \EOS [0]{\spacefactor3000\relax}%
\providecommand \BibitemShut  [1]{\csname bibitem#1\endcsname}%
\let\auto@bib@innerbib\@empty
\bibitem [{\citenamefont {Feynman}(1982)}]{Feynman1982}%
  \BibitemOpen
  \bibfield  {author} {\bibinfo {author} {\bibfnamefont {R.~P.}\ \bibnamefont
  {Feynman}},\ }\href@noop {} {\bibfield  {journal} {\bibinfo  {journal} {Int.
  J. Theor. Phys.}\ }\textbf {\bibinfo {volume} {21}},\ \bibinfo {pages} {467}
  (\bibinfo {year} {1982})}\BibitemShut {NoStop}%
\bibitem [{\citenamefont {Buluta}\ and\ \citenamefont
  {Nori}(2009)}]{Buluta2009}%
  \BibitemOpen
  \bibfield  {author} {\bibinfo {author} {\bibfnamefont {I.}~\bibnamefont
  {Buluta}}\ and\ \bibinfo {author} {\bibfnamefont {F.}~\bibnamefont {Nori}},\
  }\href@noop {} {\bibfield  {journal} {\bibinfo  {journal} {Science}\ }\textbf
  {\bibinfo {volume} {326}},\ \bibinfo {pages} {108} (\bibinfo {year}
  {2009})}\BibitemShut {NoStop}%
\bibitem [{\citenamefont {Georgescu}\ \emph {et~al.}(2014)\citenamefont
  {Georgescu}, \citenamefont {Ashhab},\ and\ \citenamefont
  {Nori}}]{Georgescu2014}%
  \BibitemOpen
  \bibfield  {author} {\bibinfo {author} {\bibfnamefont {I.~M.}\ \bibnamefont
  {Georgescu}}, \bibinfo {author} {\bibfnamefont {S.}~\bibnamefont {Ashhab}}, \
  and\ \bibinfo {author} {\bibfnamefont {F.}~\bibnamefont {Nori}},\ }\href@noop
  {} {\bibfield  {journal} {\bibinfo  {journal} {Rev. Mod. Phys.}\ }\textbf
  {\bibinfo {volume} {86}},\ \bibinfo {pages} {153} (\bibinfo {year}
  {2014})}\BibitemShut {NoStop}%
\bibitem [{\citenamefont {Du}\ \emph {et~al.}(2010)\citenamefont {Du},
  \citenamefont {Xu}, \citenamefont {Peng}, \citenamefont {Wang}, \citenamefont
  {Wu},\ and\ \citenamefont {Lu}}]{Du2010}%
  \BibitemOpen
  \bibfield  {author} {\bibinfo {author} {\bibfnamefont {J.}~\bibnamefont
  {Du}}, \bibinfo {author} {\bibfnamefont {N.}~\bibnamefont {Xu}}, \bibinfo
  {author} {\bibfnamefont {X.}~\bibnamefont {Peng}}, \bibinfo {author}
  {\bibfnamefont {P.}~\bibnamefont {Wang}}, \bibinfo {author} {\bibfnamefont
  {S.}~\bibnamefont {Wu}}, \ and\ \bibinfo {author} {\bibfnamefont
  {D.}~\bibnamefont {Lu}},\ }\href@noop {} {\bibfield  {journal} {\bibinfo
  {journal} {Phys. Rev. Lett.}\ }\textbf {\bibinfo {volume} {104}},\ \bibinfo
  {pages} {030502} (\bibinfo {year} {2010})}\BibitemShut {NoStop}%
\bibitem [{\citenamefont {Lu}\ \emph {et~al.}(2011)\citenamefont {Lu},
  \citenamefont {Xu}, \citenamefont {Xu}, \citenamefont {Chen}, \citenamefont
  {Gong}, \citenamefont {Peng},\ and\ \citenamefont {Du}}]{Lu2011}%
  \BibitemOpen
  \bibfield  {author} {\bibinfo {author} {\bibfnamefont {D.}~\bibnamefont
  {Lu}}, \bibinfo {author} {\bibfnamefont {N.}~\bibnamefont {Xu}}, \bibinfo
  {author} {\bibfnamefont {R.}~\bibnamefont {Xu}}, \bibinfo {author}
  {\bibfnamefont {H.}~\bibnamefont {Chen}}, \bibinfo {author} {\bibfnamefont
  {J.}~\bibnamefont {Gong}}, \bibinfo {author} {\bibfnamefont {X.}~\bibnamefont
  {Peng}}, \ and\ \bibinfo {author} {\bibfnamefont {J.}~\bibnamefont {Du}},\
  }\href@noop {} {\bibfield  {journal} {\bibinfo  {journal} {Phys. Rev. Lett.}\
  }\textbf {\bibinfo {volume} {107}},\ \bibinfo {pages} {020501} (\bibinfo
  {year} {2011})}\BibitemShut {NoStop}%
\bibitem [{\citenamefont {Lanyon}\ \emph {et~al.}(2010)\citenamefont {Lanyon},
  \citenamefont {Whitfield}, \citenamefont {Gillett}, \citenamefont {Goggin},
  \citenamefont {Almeida}, \citenamefont {Kassal}, \citenamefont {Biamonte},
  \citenamefont {Mohseni}, \citenamefont {Powell}, \citenamefont {Barbieri},
  \citenamefont {Aspuru-Guzik},\ and\ \citenamefont {White}}]{Lanyon2010}%
  \BibitemOpen
  \bibfield  {author} {\bibinfo {author} {\bibfnamefont {B.~P.}\ \bibnamefont
  {Lanyon}}, \bibinfo {author} {\bibfnamefont {J.~D.}\ \bibnamefont
  {Whitfield}}, \bibinfo {author} {\bibfnamefont {G.~G.}\ \bibnamefont
  {Gillett}}, \bibinfo {author} {\bibfnamefont {M.~E.}\ \bibnamefont {Goggin}},
  \bibinfo {author} {\bibfnamefont {M.~P.}\ \bibnamefont {Almeida}}, \bibinfo
  {author} {\bibfnamefont {I.}~\bibnamefont {Kassal}}, \bibinfo {author}
  {\bibfnamefont {J.~D.}\ \bibnamefont {Biamonte}}, \bibinfo {author}
  {\bibfnamefont {M.}~\bibnamefont {Mohseni}}, \bibinfo {author} {\bibfnamefont
  {B.~J.}\ \bibnamefont {Powell}}, \bibinfo {author} {\bibfnamefont
  {M.}~\bibnamefont {Barbieri}}, \bibinfo {author} {\bibfnamefont
  {A.}~\bibnamefont {Aspuru-Guzik}}, \ and\ \bibinfo {author} {\bibfnamefont
  {A.~G.}\ \bibnamefont {White}},\ }\href@noop {} {\bibfield  {journal}
  {\bibinfo  {journal} {Nat. Chem.}\ }\textbf {\bibinfo {volume} {2}},\
  \bibinfo {pages} {106} (\bibinfo {year} {2010})}\BibitemShut {NoStop}%
\bibitem [{\citenamefont {Aspuru-Guzik}\ and\ \citenamefont
  {Walther}(2012)}]{Aspuru-Guzik2012}%
  \BibitemOpen
  \bibfield  {author} {\bibinfo {author} {\bibfnamefont {A.}~\bibnamefont
  {Aspuru-Guzik}}\ and\ \bibinfo {author} {\bibfnamefont {P.}~\bibnamefont
  {Walther}},\ }\href@noop {} {\bibfield  {journal} {\bibinfo  {journal} {Nat.
  Phys.}\ }\textbf {\bibinfo {volume} {8}},\ \bibinfo {pages} {285} (\bibinfo
  {year} {2012})}\BibitemShut {NoStop}%
\bibitem [{\citenamefont {Peruzzo}\ \emph {et~al.}(2014)\citenamefont
  {Peruzzo}, \citenamefont {McClean}, \citenamefont {Shadbolt}, \citenamefont
  {Yung}, \citenamefont {Zhou}, \citenamefont {Love}, \citenamefont
  {Aspuru-Guzik},\ and\ \citenamefont {O'Brien}}]{Peruzzo2014}%
  \BibitemOpen
  \bibfield  {author} {\bibinfo {author} {\bibfnamefont {A.}~\bibnamefont
  {Peruzzo}}, \bibinfo {author} {\bibfnamefont {J.}~\bibnamefont {McClean}},
  \bibinfo {author} {\bibfnamefont {P.}~\bibnamefont {Shadbolt}}, \bibinfo
  {author} {\bibfnamefont {M.-H.}\ \bibnamefont {Yung}}, \bibinfo {author}
  {\bibfnamefont {X.-Q.}\ \bibnamefont {Zhou}}, \bibinfo {author}
  {\bibfnamefont {P.~J.}\ \bibnamefont {Love}}, \bibinfo {author}
  {\bibfnamefont {A.}~\bibnamefont {Aspuru-Guzik}}, \ and\ \bibinfo {author}
  {\bibfnamefont {J.~L.}\ \bibnamefont {O'Brien}},\ }\href@noop {} {\bibfield
  {journal} {\bibinfo  {journal} {Nat. Commun.}\ }\textbf {\bibinfo {volume}
  {5}},\ \bibinfo {pages} {4213} (\bibinfo {year} {2014})}\BibitemShut
  {NoStop}%
\bibitem [{\citenamefont {Hartmann}(2016)}]{Hartmann2016}%
  \BibitemOpen
  \bibfield  {author} {\bibinfo {author} {\bibfnamefont {M.~J.}\ \bibnamefont
  {Hartmann}},\ }\href@noop {} {\bibfield  {journal} {\bibinfo  {journal} {J.
  Opt.}\ }\textbf {\bibinfo {volume} {18}},\ \bibinfo {pages} {104005}
  (\bibinfo {year} {2016})}\BibitemShut {NoStop}%
\bibitem [{\citenamefont {Bloch}\ \emph {et~al.}(2012)\citenamefont {Bloch},
  \citenamefont {Dalibard},\ and\ \citenamefont
  {Nascimb{\`{e}}ne}}]{Bloch2012}%
  \BibitemOpen
  \bibfield  {author} {\bibinfo {author} {\bibfnamefont {I.}~\bibnamefont
  {Bloch}}, \bibinfo {author} {\bibfnamefont {J.}~\bibnamefont {Dalibard}}, \
  and\ \bibinfo {author} {\bibfnamefont {S.}~\bibnamefont {Nascimb{\`{e}}ne}},\
  }\href@noop {} {\bibfield  {journal} {\bibinfo  {journal} {Nat. Phys.}\
  }\textbf {\bibinfo {volume} {8}},\ \bibinfo {pages} {267} (\bibinfo {year}
  {2012})}\BibitemShut {NoStop}%
\bibitem [{\citenamefont {Schaetz}\ \emph {et~al.}(2013)\citenamefont
  {Schaetz}, \citenamefont {Monroe},\ and\ \citenamefont
  {Esslinger}}]{Schaetz2013}%
  \BibitemOpen
  \bibfield  {author} {\bibinfo {author} {\bibfnamefont {T.}~\bibnamefont
  {Schaetz}}, \bibinfo {author} {\bibfnamefont {C.~R.}\ \bibnamefont {Monroe}},
  \ and\ \bibinfo {author} {\bibfnamefont {T.}~\bibnamefont {Esslinger}},\
  }\href@noop {} {\bibfield  {journal} {\bibinfo  {journal} {New J. Phys.}\
  }\textbf {\bibinfo {volume} {15}},\ \bibinfo {pages} {085009} (\bibinfo
  {year} {2013})}\BibitemShut {NoStop}%
\bibitem [{\citenamefont {Blatt}\ and\ \citenamefont {Roos}(2012)}]{Blatt2012}%
  \BibitemOpen
  \bibfield  {author} {\bibinfo {author} {\bibfnamefont {R.}~\bibnamefont
  {Blatt}}\ and\ \bibinfo {author} {\bibfnamefont {C.~F.}\ \bibnamefont
  {Roos}},\ }\href@noop {} {\bibfield  {journal} {\bibinfo  {journal} {Nat.
  Phys.}\ }\textbf {\bibinfo {volume} {8}},\ \bibinfo {pages} {277} (\bibinfo
  {year} {2012})}\BibitemShut {NoStop}%
\bibitem [{\citenamefont {Zhang}\ \emph {et~al.}(2017)\citenamefont {Zhang},
  \citenamefont {Pagano}, \citenamefont {Hess}, \citenamefont {Kyprianidis},
  \citenamefont {Becker}, \citenamefont {Kaplan}, \citenamefont {Gorshkov},
  \citenamefont {Gong},\ and\ \citenamefont {Monroe}}]{Zhang2017}%
  \BibitemOpen
  \bibfield  {author} {\bibinfo {author} {\bibfnamefont {J.}~\bibnamefont
  {Zhang}}, \bibinfo {author} {\bibfnamefont {G.}~\bibnamefont {Pagano}},
  \bibinfo {author} {\bibfnamefont {P.~W.}\ \bibnamefont {Hess}}, \bibinfo
  {author} {\bibfnamefont {A.}~\bibnamefont {Kyprianidis}}, \bibinfo {author}
  {\bibfnamefont {P.}~\bibnamefont {Becker}}, \bibinfo {author} {\bibfnamefont
  {H.}~\bibnamefont {Kaplan}}, \bibinfo {author} {\bibfnamefont {A.~V.}\
  \bibnamefont {Gorshkov}}, \bibinfo {author} {\bibfnamefont {Z.-X.}\
  \bibnamefont {Gong}}, \ and\ \bibinfo {author} {\bibfnamefont
  {C.}~\bibnamefont {Monroe}},\ }\href@noop {} {\bibfield  {journal} {\bibinfo
  {journal} {Nature}\ }\textbf {\bibinfo {volume} {551}},\ \bibinfo {pages}
  {601} (\bibinfo {year} {2017})}\BibitemShut {NoStop}%
\bibitem [{\citenamefont {Schmidt}\ and\ \citenamefont
  {Koch}(2013)}]{Schmidt2013}%
  \BibitemOpen
  \bibfield  {author} {\bibinfo {author} {\bibfnamefont {S.}~\bibnamefont
  {Schmidt}}\ and\ \bibinfo {author} {\bibfnamefont {J.}~\bibnamefont {Koch}},\
  }\href@noop {} {\bibfield  {journal} {\bibinfo  {journal} {Ann. Phys.}\
  }\textbf {\bibinfo {volume} {525}},\ \bibinfo {pages} {395} (\bibinfo {year}
  {2013})}\BibitemShut {NoStop}%
\bibitem [{\citenamefont {Salath\'e}\ \emph {et~al.}(2015)\citenamefont
  {Salath\'e}, \citenamefont {Mondal}, \citenamefont {Oppliger}, \citenamefont
  {Heinsoo}, \citenamefont {Kurpiers}, \citenamefont
  {Poto\ifmmode~\check{c}\else \v{c}\fi{}nik}, \citenamefont {Mezzacapo},
  \citenamefont {Las~Heras}, \citenamefont {Lamata}, \citenamefont {Solano},
  \citenamefont {Filipp},\ and\ \citenamefont {Wallraff}}]{Salathe2015}%
  \BibitemOpen
  \bibfield  {author} {\bibinfo {author} {\bibfnamefont {Y.}~\bibnamefont
  {Salath\'e}}, \bibinfo {author} {\bibfnamefont {M.}~\bibnamefont {Mondal}},
  \bibinfo {author} {\bibfnamefont {M.}~\bibnamefont {Oppliger}}, \bibinfo
  {author} {\bibfnamefont {J.}~\bibnamefont {Heinsoo}}, \bibinfo {author}
  {\bibfnamefont {P.}~\bibnamefont {Kurpiers}}, \bibinfo {author}
  {\bibfnamefont {A.}~\bibnamefont {Poto\ifmmode~\check{c}\else
  \v{c}\fi{}nik}}, \bibinfo {author} {\bibfnamefont {A.}~\bibnamefont
  {Mezzacapo}}, \bibinfo {author} {\bibfnamefont {U.}~\bibnamefont
  {Las~Heras}}, \bibinfo {author} {\bibfnamefont {L.}~\bibnamefont {Lamata}},
  \bibinfo {author} {\bibfnamefont {E.}~\bibnamefont {Solano}}, \bibinfo
  {author} {\bibfnamefont {S.}~\bibnamefont {Filipp}}, \ and\ \bibinfo {author}
  {\bibfnamefont {A.}~\bibnamefont {Wallraff}},\ }\href@noop {} {\bibfield
  {journal} {\bibinfo  {journal} {Phys. Rev. X}\ }\textbf {\bibinfo {volume}
  {5}},\ \bibinfo {pages} {021027} (\bibinfo {year} {2015})}\BibitemShut
  {NoStop}%
\bibitem [{\citenamefont {Barends}\ \emph {et~al.}(2016)\citenamefont
  {Barends}, \citenamefont {Shabani}, \citenamefont {Lamata}, \citenamefont
  {Kelly}, \citenamefont {Mezzacapo}, \citenamefont {Heras}, \citenamefont
  {Babbush}, \citenamefont {Fowler}, \citenamefont {Campbell}, \citenamefont
  {Chen}, \citenamefont {Chen}, \citenamefont {Chiaro}, \citenamefont
  {Dunsworth}, \citenamefont {Jeffrey}, \citenamefont {Lucero}, \citenamefont
  {Megrant}, \citenamefont {Mutus}, \citenamefont {Neeley}, \citenamefont
  {Neill}, \citenamefont {O'Malley}, \citenamefont {Quintana}, \citenamefont
  {Roushan}, \citenamefont {Sank}, \citenamefont {Vainsencher}, \citenamefont
  {Wenner}, \citenamefont {White}, \citenamefont {Solano}, \citenamefont
  {Neven},\ and\ \citenamefont {Martinis}}]{Barends2016}%
  \BibitemOpen
  \bibfield  {author} {\bibinfo {author} {\bibfnamefont {R.}~\bibnamefont
  {Barends}}, \bibinfo {author} {\bibfnamefont {A.}~\bibnamefont {Shabani}},
  \bibinfo {author} {\bibfnamefont {L.}~\bibnamefont {Lamata}}, \bibinfo
  {author} {\bibfnamefont {J.}~\bibnamefont {Kelly}}, \bibinfo {author}
  {\bibfnamefont {A.}~\bibnamefont {Mezzacapo}}, \bibinfo {author}
  {\bibfnamefont {U.~L.}\ \bibnamefont {Heras}}, \bibinfo {author}
  {\bibfnamefont {R.}~\bibnamefont {Babbush}}, \bibinfo {author} {\bibfnamefont
  {A.~G.}\ \bibnamefont {Fowler}}, \bibinfo {author} {\bibfnamefont
  {B.}~\bibnamefont {Campbell}}, \bibinfo {author} {\bibfnamefont
  {Y.}~\bibnamefont {Chen}}, \bibinfo {author} {\bibfnamefont {Z.}~\bibnamefont
  {Chen}}, \bibinfo {author} {\bibfnamefont {B.}~\bibnamefont {Chiaro}},
  \bibinfo {author} {\bibfnamefont {A.}~\bibnamefont {Dunsworth}}, \bibinfo
  {author} {\bibfnamefont {E.}~\bibnamefont {Jeffrey}}, \bibinfo {author}
  {\bibfnamefont {E.}~\bibnamefont {Lucero}}, \bibinfo {author} {\bibfnamefont
  {A.}~\bibnamefont {Megrant}}, \bibinfo {author} {\bibfnamefont {J.~Y.}\
  \bibnamefont {Mutus}}, \bibinfo {author} {\bibfnamefont {M.}~\bibnamefont
  {Neeley}}, \bibinfo {author} {\bibfnamefont {C.}~\bibnamefont {Neill}},
  \bibinfo {author} {\bibfnamefont {P.~J.~J.}\ \bibnamefont {O'Malley}},
  \bibinfo {author} {\bibfnamefont {C.}~\bibnamefont {Quintana}}, \bibinfo
  {author} {\bibfnamefont {P.}~\bibnamefont {Roushan}}, \bibinfo {author}
  {\bibfnamefont {D.}~\bibnamefont {Sank}}, \bibinfo {author} {\bibfnamefont
  {A.}~\bibnamefont {Vainsencher}}, \bibinfo {author} {\bibfnamefont
  {J.}~\bibnamefont {Wenner}}, \bibinfo {author} {\bibfnamefont {T.~C.}\
  \bibnamefont {White}}, \bibinfo {author} {\bibfnamefont {E.}~\bibnamefont
  {Solano}}, \bibinfo {author} {\bibfnamefont {H.}~\bibnamefont {Neven}}, \
  and\ \bibinfo {author} {\bibfnamefont {J.~M.}\ \bibnamefont {Martinis}},\
  }\href@noop {} {\bibfield  {journal} {\bibinfo  {journal} {Nature}\ }\textbf
  {\bibinfo {volume} {534}},\ \bibinfo {pages} {222} (\bibinfo {year}
  {2016})}\BibitemShut {NoStop}%
\bibitem [{\citenamefont {Langford}\ \emph {et~al.}(2017)\citenamefont
  {Langford}, \citenamefont {Sagastizabal}, \citenamefont {Kounalakis},
  \citenamefont {Dickel}, \citenamefont {Bruno}, \citenamefont {Luthi},
  \citenamefont {Thoen}, \citenamefont {Endo},\ and\ \citenamefont
  {DiCarlo}}]{Langford2017}%
  \BibitemOpen
  \bibfield  {author} {\bibinfo {author} {\bibfnamefont {N.~K.}\ \bibnamefont
  {Langford}}, \bibinfo {author} {\bibfnamefont {R.}~\bibnamefont
  {Sagastizabal}}, \bibinfo {author} {\bibfnamefont {M.}~\bibnamefont
  {Kounalakis}}, \bibinfo {author} {\bibfnamefont {C.}~\bibnamefont {Dickel}},
  \bibinfo {author} {\bibfnamefont {A.}~\bibnamefont {Bruno}}, \bibinfo
  {author} {\bibfnamefont {F.}~\bibnamefont {Luthi}}, \bibinfo {author}
  {\bibfnamefont {D.~J.}\ \bibnamefont {Thoen}}, \bibinfo {author}
  {\bibfnamefont {A.}~\bibnamefont {Endo}}, \ and\ \bibinfo {author}
  {\bibfnamefont {L.}~\bibnamefont {DiCarlo}},\ }\href@noop {} {\bibfield
  {journal} {\bibinfo  {journal} {Nat. Commun.}\ }\textbf {\bibinfo {volume}
  {8}},\ \bibinfo {pages} {1715} (\bibinfo {year} {2017})}\BibitemShut
  {NoStop}%
\bibitem [{\citenamefont {O'Malley}\ \emph {et~al.}(2016)\citenamefont
  {O'Malley}, \citenamefont {Babbush}, \citenamefont {Kivlichan}, \citenamefont
  {Romero}, \citenamefont {McClean}, \citenamefont {Barends}, \citenamefont
  {Kelly}, \citenamefont {Roushan}, \citenamefont {Tranter}, \citenamefont
  {Ding}, \citenamefont {Campbell}, \citenamefont {Chen}, \citenamefont {Chen},
  \citenamefont {Chiaro}, \citenamefont {Dunsworth}, \citenamefont {Fowler},
  \citenamefont {Jeffrey}, \citenamefont {Lucero}, \citenamefont {Megrant},
  \citenamefont {Mutus}, \citenamefont {Neeley}, \citenamefont {Neill},
  \citenamefont {Quintana}, \citenamefont {Sank}, \citenamefont {Vainsencher},
  \citenamefont {Wenner}, \citenamefont {White}, \citenamefont {Coveney},
  \citenamefont {Love}, \citenamefont {Neven}, \citenamefont {Aspuru-Guzik},\
  and\ \citenamefont {Martinis}}]{OMalley2016}%
  \BibitemOpen
  \bibfield  {author} {\bibinfo {author} {\bibfnamefont {P.~J.~J.}\
  \bibnamefont {O'Malley}}, \bibinfo {author} {\bibfnamefont {R.}~\bibnamefont
  {Babbush}}, \bibinfo {author} {\bibfnamefont {I.~D.}\ \bibnamefont
  {Kivlichan}}, \bibinfo {author} {\bibfnamefont {J.}~\bibnamefont {Romero}},
  \bibinfo {author} {\bibfnamefont {J.~R.}\ \bibnamefont {McClean}}, \bibinfo
  {author} {\bibfnamefont {R.}~\bibnamefont {Barends}}, \bibinfo {author}
  {\bibfnamefont {J.}~\bibnamefont {Kelly}}, \bibinfo {author} {\bibfnamefont
  {P.}~\bibnamefont {Roushan}}, \bibinfo {author} {\bibfnamefont
  {A.}~\bibnamefont {Tranter}}, \bibinfo {author} {\bibfnamefont
  {N.}~\bibnamefont {Ding}}, \bibinfo {author} {\bibfnamefont {B.}~\bibnamefont
  {Campbell}}, \bibinfo {author} {\bibfnamefont {Y.}~\bibnamefont {Chen}},
  \bibinfo {author} {\bibfnamefont {Z.}~\bibnamefont {Chen}}, \bibinfo {author}
  {\bibfnamefont {B.}~\bibnamefont {Chiaro}}, \bibinfo {author} {\bibfnamefont
  {A.}~\bibnamefont {Dunsworth}}, \bibinfo {author} {\bibfnamefont {A.~G.}\
  \bibnamefont {Fowler}}, \bibinfo {author} {\bibfnamefont {E.}~\bibnamefont
  {Jeffrey}}, \bibinfo {author} {\bibfnamefont {E.}~\bibnamefont {Lucero}},
  \bibinfo {author} {\bibfnamefont {A.}~\bibnamefont {Megrant}}, \bibinfo
  {author} {\bibfnamefont {J.~Y.}\ \bibnamefont {Mutus}}, \bibinfo {author}
  {\bibfnamefont {M.}~\bibnamefont {Neeley}}, \bibinfo {author} {\bibfnamefont
  {C.}~\bibnamefont {Neill}}, \bibinfo {author} {\bibfnamefont
  {C.}~\bibnamefont {Quintana}}, \bibinfo {author} {\bibfnamefont
  {D.}~\bibnamefont {Sank}}, \bibinfo {author} {\bibfnamefont {A.}~\bibnamefont
  {Vainsencher}}, \bibinfo {author} {\bibfnamefont {J.}~\bibnamefont {Wenner}},
  \bibinfo {author} {\bibfnamefont {T.~C.}\ \bibnamefont {White}}, \bibinfo
  {author} {\bibfnamefont {P.~V.}\ \bibnamefont {Coveney}}, \bibinfo {author}
  {\bibfnamefont {P.~J.}\ \bibnamefont {Love}}, \bibinfo {author}
  {\bibfnamefont {H.}~\bibnamefont {Neven}}, \bibinfo {author} {\bibfnamefont
  {A.}~\bibnamefont {Aspuru-Guzik}}, \ and\ \bibinfo {author} {\bibfnamefont
  {J.~M.}\ \bibnamefont {Martinis}},\ }\href@noop {} {\bibfield  {journal}
  {\bibinfo  {journal} {Phys. Rev. X}\ }\textbf {\bibinfo {volume} {6}},\
  \bibinfo {pages} {031007} (\bibinfo {year} {2016})}\BibitemShut {NoStop}%
\bibitem [{\citenamefont {Roushan}\ \emph {et~al.}(2016)\citenamefont
  {Roushan}, \citenamefont {Neill}, \citenamefont {Megrant}, \citenamefont
  {Chen}, \citenamefont {Babbush}, \citenamefont {Barends}, \citenamefont
  {Campbell}, \citenamefont {Chen}, \citenamefont {Chiaro}, \citenamefont
  {Dunsworth}, \citenamefont {Fowler}, \citenamefont {Jeffrey}, \citenamefont
  {Kelly}, \citenamefont {Lucero}, \citenamefont {Mutus}, \citenamefont
  {O'Malley}, \citenamefont {Neeley}, \citenamefont {Quintana}, \citenamefont
  {Sank}, \citenamefont {Vainsencher}, \citenamefont {Wenner}, \citenamefont
  {White}, \citenamefont {Kapit}, \citenamefont {Neven},\ and\ \citenamefont
  {Martinis}}]{Roushan2016}%
  \BibitemOpen
  \bibfield  {author} {\bibinfo {author} {\bibfnamefont {P.}~\bibnamefont
  {Roushan}}, \bibinfo {author} {\bibfnamefont {C.}~\bibnamefont {Neill}},
  \bibinfo {author} {\bibfnamefont {A.}~\bibnamefont {Megrant}}, \bibinfo
  {author} {\bibfnamefont {Y.}~\bibnamefont {Chen}}, \bibinfo {author}
  {\bibfnamefont {R.}~\bibnamefont {Babbush}}, \bibinfo {author} {\bibfnamefont
  {R.}~\bibnamefont {Barends}}, \bibinfo {author} {\bibfnamefont
  {B.}~\bibnamefont {Campbell}}, \bibinfo {author} {\bibfnamefont
  {Z.}~\bibnamefont {Chen}}, \bibinfo {author} {\bibfnamefont {B.}~\bibnamefont
  {Chiaro}}, \bibinfo {author} {\bibfnamefont {A.}~\bibnamefont {Dunsworth}},
  \bibinfo {author} {\bibfnamefont {A.}~\bibnamefont {Fowler}}, \bibinfo
  {author} {\bibfnamefont {E.}~\bibnamefont {Jeffrey}}, \bibinfo {author}
  {\bibfnamefont {J.}~\bibnamefont {Kelly}}, \bibinfo {author} {\bibfnamefont
  {E.}~\bibnamefont {Lucero}}, \bibinfo {author} {\bibfnamefont
  {J.}~\bibnamefont {Mutus}}, \bibinfo {author} {\bibfnamefont {P.~J.~J.}\
  \bibnamefont {O'Malley}}, \bibinfo {author} {\bibfnamefont {M.}~\bibnamefont
  {Neeley}}, \bibinfo {author} {\bibfnamefont {C.}~\bibnamefont {Quintana}},
  \bibinfo {author} {\bibfnamefont {D.}~\bibnamefont {Sank}}, \bibinfo {author}
  {\bibfnamefont {A.}~\bibnamefont {Vainsencher}}, \bibinfo {author}
  {\bibfnamefont {J.}~\bibnamefont {Wenner}}, \bibinfo {author} {\bibfnamefont
  {T.}~\bibnamefont {White}}, \bibinfo {author} {\bibfnamefont
  {E.}~\bibnamefont {Kapit}}, \bibinfo {author} {\bibfnamefont
  {H.}~\bibnamefont {Neven}}, \ and\ \bibinfo {author} {\bibfnamefont
  {J.}~\bibnamefont {Martinis}},\ }\href@noop {} {\bibfield  {journal}
  {\bibinfo  {journal} {Nat. Phys.}\ }\textbf {\bibinfo {volume} {13}},\
  \bibinfo {pages} {146} (\bibinfo {year} {2016})}\BibitemShut {NoStop}%
\bibitem [{\citenamefont {Wendin}(2017)}]{Wendin2017}%
  \BibitemOpen
  \bibfield  {author} {\bibinfo {author} {\bibfnamefont {G.}~\bibnamefont
  {Wendin}},\ }\href@noop {} {\bibfield  {journal} {\bibinfo  {journal} {Rep.
  Prog. Phys.}\ }\textbf {\bibinfo {volume} {80}},\ \bibinfo {pages} {106001}
  (\bibinfo {year} {2017})}\BibitemShut {NoStop}%
\bibitem [{\citenamefont {Heras}\ \emph {et~al.}(2014)\citenamefont {Heras},
  \citenamefont {Mezzacapo}, \citenamefont {Lamata}, \citenamefont {Filipp},
  \citenamefont {Wallraff},\ and\ \citenamefont {Solano}}]{Heras2014}%
  \BibitemOpen
  \bibfield  {author} {\bibinfo {author} {\bibfnamefont {U.~L.}\ \bibnamefont
  {Heras}}, \bibinfo {author} {\bibfnamefont {A.}~\bibnamefont {Mezzacapo}},
  \bibinfo {author} {\bibfnamefont {L.}~\bibnamefont {Lamata}}, \bibinfo
  {author} {\bibfnamefont {S.}~\bibnamefont {Filipp}}, \bibinfo {author}
  {\bibfnamefont {A.}~\bibnamefont {Wallraff}}, \ and\ \bibinfo {author}
  {\bibfnamefont {E.}~\bibnamefont {Solano}},\ }\href@noop {} {\bibfield
  {journal} {\bibinfo  {journal} {Phys. Rev. Lett.}\ }\textbf {\bibinfo
  {volume} {112}},\ \bibinfo {pages} {200501} (\bibinfo {year}
  {2014})}\BibitemShut {NoStop}%
\bibitem [{\citenamefont {Barends}\ \emph {et~al.}(2015)\citenamefont
  {Barends}, \citenamefont {Lamata}, \citenamefont {Kelly}, \citenamefont
  {Garc{\'{i}}a-{\'{A}}lvarez}, \citenamefont {Fowler}, \citenamefont
  {Megrant}, \citenamefont {Jeffrey}, \citenamefont {White}, \citenamefont
  {Sank}, \citenamefont {Mutus}, \citenamefont {Campbell}, \citenamefont
  {Chen}, \citenamefont {Chen}, \citenamefont {Chiaro}, \citenamefont
  {Dunsworth}, \citenamefont {Hoi}, \citenamefont {Neill}, \citenamefont
  {O'Malley}, \citenamefont {Quintana}, \citenamefont {Roushan}, \citenamefont
  {Vainsencher}, \citenamefont {Wenner}, \citenamefont {Solano},\ and\
  \citenamefont {Martinis}}]{Barends2015}%
  \BibitemOpen
  \bibfield  {author} {\bibinfo {author} {\bibfnamefont {R.}~\bibnamefont
  {Barends}}, \bibinfo {author} {\bibfnamefont {L.}~\bibnamefont {Lamata}},
  \bibinfo {author} {\bibfnamefont {J.}~\bibnamefont {Kelly}}, \bibinfo
  {author} {\bibfnamefont {L.}~\bibnamefont {Garc{\'{i}}a-{\'{A}}lvarez}},
  \bibinfo {author} {\bibfnamefont {A.~G.}\ \bibnamefont {Fowler}}, \bibinfo
  {author} {\bibfnamefont {A.}~\bibnamefont {Megrant}}, \bibinfo {author}
  {\bibfnamefont {E.}~\bibnamefont {Jeffrey}}, \bibinfo {author} {\bibfnamefont
  {T.~C.}\ \bibnamefont {White}}, \bibinfo {author} {\bibfnamefont
  {D.}~\bibnamefont {Sank}}, \bibinfo {author} {\bibfnamefont {J.~Y.}\
  \bibnamefont {Mutus}}, \bibinfo {author} {\bibfnamefont {B.}~\bibnamefont
  {Campbell}}, \bibinfo {author} {\bibfnamefont {Y.}~\bibnamefont {Chen}},
  \bibinfo {author} {\bibfnamefont {Z.}~\bibnamefont {Chen}}, \bibinfo {author}
  {\bibfnamefont {B.}~\bibnamefont {Chiaro}}, \bibinfo {author} {\bibfnamefont
  {A.}~\bibnamefont {Dunsworth}}, \bibinfo {author} {\bibfnamefont {I.-C.}\
  \bibnamefont {Hoi}}, \bibinfo {author} {\bibfnamefont {C.}~\bibnamefont
  {Neill}}, \bibinfo {author} {\bibfnamefont {P.~J.~J.}\ \bibnamefont
  {O'Malley}}, \bibinfo {author} {\bibfnamefont {C.}~\bibnamefont {Quintana}},
  \bibinfo {author} {\bibfnamefont {P.}~\bibnamefont {Roushan}}, \bibinfo
  {author} {\bibfnamefont {A.}~\bibnamefont {Vainsencher}}, \bibinfo {author}
  {\bibfnamefont {J.}~\bibnamefont {Wenner}}, \bibinfo {author} {\bibfnamefont
  {E.}~\bibnamefont {Solano}}, \ and\ \bibinfo {author} {\bibfnamefont {J.~M.}\
  \bibnamefont {Martinis}},\ }\href@noop {} {\bibfield  {journal} {\bibinfo
  {journal} {Nat. Commun.}\ }\textbf {\bibinfo {volume} {6}},\ \bibinfo {pages}
  {7654} (\bibinfo {year} {2015})}\BibitemShut {NoStop}%
\bibitem [{\citenamefont {Suzuki}(1990)}]{Suzuki1990}%
  \BibitemOpen
  \bibfield  {author} {\bibinfo {author} {\bibfnamefont {M.}~\bibnamefont
  {Suzuki}},\ }\href@noop {} {\bibfield  {journal} {\bibinfo  {journal} {Phys.
  Lett.}\ }\textbf {\bibinfo {volume} {146}},\ \bibinfo {pages} {319 }
  (\bibinfo {year} {1990})}\BibitemShut {NoStop}%
\bibitem [{\citenamefont {Kandala}\ \emph {et~al.}(2017)\citenamefont
  {Kandala}, \citenamefont {Mezzacapo}, \citenamefont {Temme}, \citenamefont
  {Takita}, \citenamefont {Brink}, \citenamefont {Chow},\ and\ \citenamefont
  {Gambetta}}]{Kandala2017}%
  \BibitemOpen
  \bibfield  {author} {\bibinfo {author} {\bibfnamefont {A.}~\bibnamefont
  {Kandala}}, \bibinfo {author} {\bibfnamefont {A.}~\bibnamefont {Mezzacapo}},
  \bibinfo {author} {\bibfnamefont {K.}~\bibnamefont {Temme}}, \bibinfo
  {author} {\bibfnamefont {M.}~\bibnamefont {Takita}}, \bibinfo {author}
  {\bibfnamefont {M.}~\bibnamefont {Brink}}, \bibinfo {author} {\bibfnamefont
  {J.~M.}\ \bibnamefont {Chow}}, \ and\ \bibinfo {author} {\bibfnamefont
  {J.~M.}\ \bibnamefont {Gambetta}},\ }\href@noop {} {\bibfield  {journal}
  {\bibinfo  {journal} {Nature}\ }\textbf {\bibinfo {volume} {549}},\ \bibinfo
  {pages} {242} (\bibinfo {year} {2017})}\BibitemShut {NoStop}%
\bibitem [{\citenamefont {Farhi}\ \emph {et~al.}(2000)\citenamefont {Farhi},
  \citenamefont {Goldstone}, \citenamefont {Gutmann},\ and\ \citenamefont
  {Sipser}}]{Farhi2000}%
  \BibitemOpen
  \bibfield  {author} {\bibinfo {author} {\bibfnamefont {E.}~\bibnamefont
  {Farhi}}, \bibinfo {author} {\bibfnamefont {J.}~\bibnamefont {Goldstone}},
  \bibinfo {author} {\bibfnamefont {S.}~\bibnamefont {Gutmann}}, \ and\
  \bibinfo {author} {\bibfnamefont {M.}~\bibnamefont {Sipser}},\ }\href@noop {}
  {\  (\bibinfo {year} {2000})},\ \bibinfo {note}
  {quant-ph/0001106}\BibitemShut {NoStop}%
\bibitem [{\citenamefont {Kassal}\ \emph {et~al.}(2011)\citenamefont {Kassal},
  \citenamefont {Whitfield}, \citenamefont {Perdomo-Ortiz}, \citenamefont
  {Yung},\ and\ \citenamefont {Aspuru-Guzik}}]{Ivan2011}%
  \BibitemOpen
  \bibfield  {author} {\bibinfo {author} {\bibfnamefont {I.}~\bibnamefont
  {Kassal}}, \bibinfo {author} {\bibfnamefont {J.~D.}\ \bibnamefont
  {Whitfield}}, \bibinfo {author} {\bibfnamefont {A.}~\bibnamefont
  {Perdomo-Ortiz}}, \bibinfo {author} {\bibfnamefont {M.-H.}\ \bibnamefont
  {Yung}}, \ and\ \bibinfo {author} {\bibfnamefont {A.}~\bibnamefont
  {Aspuru-Guzik}},\ }\href@noop {} {\bibfield  {journal} {\bibinfo  {journal}
  {Annu. Rev. Phys. Chem.}\ }\textbf {\bibinfo {volume} {62}},\ \bibinfo
  {pages} {185} (\bibinfo {year} {2011})}\BibitemShut {NoStop}%
\bibitem [{\citenamefont {Egger}\ and\ \citenamefont
  {Wilhelm}(2013)}]{Egger2013}%
  \BibitemOpen
  \bibfield  {author} {\bibinfo {author} {\bibfnamefont {D.~J.}\ \bibnamefont
  {Egger}}\ and\ \bibinfo {author} {\bibfnamefont {F.~K.}\ \bibnamefont
  {Wilhelm}},\ }\href@noop {} {\bibfield  {journal} {\bibinfo  {journal} {Phys.
  Rev. Lett.}\ }\textbf {\bibinfo {volume} {111}},\ \bibinfo {pages} {163601}
  (\bibinfo {year} {2013})}\BibitemShut {NoStop}%
\bibitem [{\citenamefont {Babbush}\ \emph {et~al.}(2014)\citenamefont
  {Babbush}, \citenamefont {Love},\ and\ \citenamefont
  {Aspuru-Guzik}}]{Babbush2014}%
  \BibitemOpen
  \bibfield  {author} {\bibinfo {author} {\bibfnamefont {R.}~\bibnamefont
  {Babbush}}, \bibinfo {author} {\bibfnamefont {P.~J.}\ \bibnamefont {Love}}, \
  and\ \bibinfo {author} {\bibfnamefont {A.}~\bibnamefont {Aspuru-Guzik}},\
  }\href@noop {} {\bibfield  {journal} {\bibinfo  {journal} {Sci. Rep.}\
  }\textbf {\bibinfo {volume} {4}},\ \bibinfo {pages} {6603} (\bibinfo {year}
  {2014})}\BibitemShut {NoStop}%
\bibitem [{\citenamefont {Kyriienko}\ and\ \citenamefont
  {S\o{}rensen}(2018)}]{Kyriienko2018}%
  \BibitemOpen
  \bibfield  {author} {\bibinfo {author} {\bibfnamefont {O.}~\bibnamefont
  {Kyriienko}}\ and\ \bibinfo {author} {\bibfnamefont {A.~S.}\ \bibnamefont
  {S\o{}rensen}},\ }\href@noop {} {\bibfield  {journal} {\bibinfo  {journal}
  {Phys. Rev. Applied}\ }\textbf {\bibinfo {volume} {9}},\ \bibinfo {pages}
  {064029} (\bibinfo {year} {2018})}\BibitemShut {NoStop}%
\bibitem [{\citenamefont {Braum{\"{u}}ller}\ \emph {et~al.}(2017)\citenamefont
  {Braum{\"{u}}ller}, \citenamefont {Marthaler}, \citenamefont {Schneider},
  \citenamefont {Stehli}, \citenamefont {Rotzinger}, \citenamefont {Weides},\
  and\ \citenamefont {Ustinov}}]{Braumuller2017}%
  \BibitemOpen
  \bibfield  {author} {\bibinfo {author} {\bibfnamefont {J.}~\bibnamefont
  {Braum{\"{u}}ller}}, \bibinfo {author} {\bibfnamefont {M.}~\bibnamefont
  {Marthaler}}, \bibinfo {author} {\bibfnamefont {A.}~\bibnamefont
  {Schneider}}, \bibinfo {author} {\bibfnamefont {A.}~\bibnamefont {Stehli}},
  \bibinfo {author} {\bibfnamefont {H.}~\bibnamefont {Rotzinger}}, \bibinfo
  {author} {\bibfnamefont {M.}~\bibnamefont {Weides}}, \ and\ \bibinfo {author}
  {\bibfnamefont {A.~V.}\ \bibnamefont {Ustinov}},\ }\href@noop {} {\bibfield
  {journal} {\bibinfo  {journal} {Nat. Commun.}\ }\textbf {\bibinfo {volume}
  {8}},\ \bibinfo {pages} {779} (\bibinfo {year} {2017})}\BibitemShut {NoStop}%
\bibitem [{\citenamefont {Schiff}(1955)}]{Schiff1955}%
  \BibitemOpen
  \bibfield  {author} {\bibinfo {author} {\bibfnamefont {L.~I.}\ \bibnamefont
  {Schiff}},\ }\href@noop {} {\emph {\bibinfo {title} {Quantum Mechanics}}}\
  (\bibinfo  {publisher} {McGraw-Hill, Singapore},\ \bibinfo {year}
  {1955})\BibitemShut {NoStop}%
\bibitem [{\citenamefont {Jordan}\ and\ \citenamefont
  {Wigner}(1928)}]{Jordan1928}%
  \BibitemOpen
  \bibfield  {author} {\bibinfo {author} {\bibfnamefont {P.}~\bibnamefont
  {Jordan}}\ and\ \bibinfo {author} {\bibfnamefont {E.}~\bibnamefont
  {Wigner}},\ }\href@noop {} {\bibfield  {journal} {\bibinfo  {journal} {Z.
  Phys.}\ }\textbf {\bibinfo {volume} {47}},\ \bibinfo {pages} {631} (\bibinfo
  {year} {1928})}\BibitemShut {NoStop}%
\bibitem [{\citenamefont {Bravyi}\ and\ \citenamefont
  {Kitaev}(2002)}]{Bravyi2002}%
  \BibitemOpen
  \bibfield  {author} {\bibinfo {author} {\bibfnamefont {S.~B.}\ \bibnamefont
  {Bravyi}}\ and\ \bibinfo {author} {\bibfnamefont {A.~Y.}\ \bibnamefont
  {Kitaev}},\ }\href@noop {} {\bibfield  {journal} {\bibinfo  {journal} {Ann.
  Phys. (N.Y.)}\ }\textbf {\bibinfo {volume} {298}},\ \bibinfo {pages} {210 }
  (\bibinfo {year} {2002})}\BibitemShut {NoStop}%
\bibitem [{\citenamefont {Cao}\ \emph {et~al.}(2015)\citenamefont {Cao},
  \citenamefont {Babbush}, \citenamefont {Biamonte},\ and\ \citenamefont
  {Kais}}]{Cao2015}%
  \BibitemOpen
  \bibfield  {author} {\bibinfo {author} {\bibfnamefont {Y.}~\bibnamefont
  {Cao}}, \bibinfo {author} {\bibfnamefont {R.}~\bibnamefont {Babbush}},
  \bibinfo {author} {\bibfnamefont {J.}~\bibnamefont {Biamonte}}, \ and\
  \bibinfo {author} {\bibfnamefont {S.}~\bibnamefont {Kais}},\ }\href@noop {}
  {\bibfield  {journal} {\bibinfo  {journal} {Phys. Rev. A}\ }\textbf {\bibinfo
  {volume} {91}},\ \bibinfo {pages} {012315} (\bibinfo {year}
  {2015})}\BibitemShut {NoStop}%
\bibitem [{\citenamefont {Lechner}\ \emph {et~al.}(2015)\citenamefont
  {Lechner}, \citenamefont {Hauke},\ and\ \citenamefont
  {Zoller}}]{Lechnere1500838}%
  \BibitemOpen
  \bibfield  {author} {\bibinfo {author} {\bibfnamefont {W.}~\bibnamefont
  {Lechner}}, \bibinfo {author} {\bibfnamefont {P.}~\bibnamefont {Hauke}}, \
  and\ \bibinfo {author} {\bibfnamefont {P.}~\bibnamefont {Zoller}},\
  }\href@noop {} {\bibfield  {journal} {\bibinfo  {journal} {Sci. Adv.}\
  }\textbf {\bibinfo {volume} {1}} (\bibinfo {year} {2015})}\BibitemShut
  {NoStop}%
\bibitem [{\citenamefont {McKay}\ \emph {et~al.}(2016)\citenamefont {McKay},
  \citenamefont {Filipp}, \citenamefont {Mezzacapo}, \citenamefont {Magesan},
  \citenamefont {Chow},\ and\ \citenamefont {Gambetta}}]{McKay2016}%
  \BibitemOpen
  \bibfield  {author} {\bibinfo {author} {\bibfnamefont {D.~C.}\ \bibnamefont
  {McKay}}, \bibinfo {author} {\bibfnamefont {S.}~\bibnamefont {Filipp}},
  \bibinfo {author} {\bibfnamefont {A.}~\bibnamefont {Mezzacapo}}, \bibinfo
  {author} {\bibfnamefont {E.}~\bibnamefont {Magesan}}, \bibinfo {author}
  {\bibfnamefont {J.~M.}\ \bibnamefont {Chow}}, \ and\ \bibinfo {author}
  {\bibfnamefont {J.~M.}\ \bibnamefont {Gambetta}},\ }\href@noop {} {\bibfield
  {journal} {\bibinfo  {journal} {Phys. Rev. Applied}\ }\textbf {\bibinfo
  {volume} {6}},\ \bibinfo {pages} {064007} (\bibinfo {year}
  {2016})}\BibitemShut {NoStop}%
\bibitem [{\citenamefont {Roth}\ \emph {et~al.}(2017)\citenamefont {Roth},
  \citenamefont {Ganzhorn}, \citenamefont {Moll}, \citenamefont {Filipp},
  \citenamefont {Salis},\ and\ \citenamefont {Schmidt}}]{Roth2017}%
  \BibitemOpen
  \bibfield  {author} {\bibinfo {author} {\bibfnamefont {M.}~\bibnamefont
  {Roth}}, \bibinfo {author} {\bibfnamefont {M.}~\bibnamefont {Ganzhorn}},
  \bibinfo {author} {\bibfnamefont {N.}~\bibnamefont {Moll}}, \bibinfo {author}
  {\bibfnamefont {S.}~\bibnamefont {Filipp}}, \bibinfo {author} {\bibfnamefont
  {G.}~\bibnamefont {Salis}}, \ and\ \bibinfo {author} {\bibfnamefont
  {S.}~\bibnamefont {Schmidt}},\ }\href@noop {} {\bibfield  {journal} {\bibinfo
   {journal} {Phys. Rev. A}\ }\textbf {\bibinfo {volume} {96}},\ \bibinfo
  {pages} {062323} (\bibinfo {year} {2017})}\BibitemShut {NoStop}%
\bibitem [{\citenamefont {Didier}\ \emph {et~al.}(2018)\citenamefont {Didier},
  \citenamefont {Sete}, \citenamefont {da~Silva},\ and\ \citenamefont
  {Rigetti}}]{Didier2018}%
  \BibitemOpen
  \bibfield  {author} {\bibinfo {author} {\bibfnamefont {N.}~\bibnamefont
  {Didier}}, \bibinfo {author} {\bibfnamefont {E.~A.}\ \bibnamefont {Sete}},
  \bibinfo {author} {\bibfnamefont {M.~P.}\ \bibnamefont {da~Silva}}, \ and\
  \bibinfo {author} {\bibfnamefont {C.}~\bibnamefont {Rigetti}},\ }\href@noop
  {} {\bibfield  {journal} {\bibinfo  {journal} {Phys. Rev. A}\ }\textbf
  {\bibinfo {volume} {97}},\ \bibinfo {pages} {022330} (\bibinfo {year}
  {2018})}\BibitemShut {NoStop}%
\bibitem [{\citenamefont {Reagor}\ \emph
  {et~al.}(2018{\natexlab{a}})\citenamefont {Reagor}, \citenamefont {Osborn},
  \citenamefont {Tezak}, \citenamefont {Staley}, \citenamefont
  {Prawiroatmodjo}, \citenamefont {Scheer}, \citenamefont {Alidoust},
  \citenamefont {Sete}, \citenamefont {Didier}, \citenamefont {da~Silva},
  \citenamefont {Acala}, \citenamefont {Angeles}, \citenamefont {Bestwick},
  \citenamefont {Block}, \citenamefont {Bloom}, \citenamefont {Bradley},
  \citenamefont {Bui}, \citenamefont {Caldwell}, \citenamefont {Capelluto},
  \citenamefont {Chilcott}, \citenamefont {Cordova}, \citenamefont {Crossman},
  \citenamefont {Curtis}, \citenamefont {Deshpande}, \citenamefont
  {El~Bouayadi}, \citenamefont {Girshovich}, \citenamefont {Hong},
  \citenamefont {Hudson}, \citenamefont {Karalekas}, \citenamefont {Kuang},
  \citenamefont {Lenihan}, \citenamefont {Manenti}, \citenamefont {Manning},
  \citenamefont {Marshall}, \citenamefont {Mohan}, \citenamefont
  {O{\textquoteright}Brien}, \citenamefont {Otterbach}, \citenamefont
  {Papageorge}, \citenamefont {Paquette}, \citenamefont {Pelstring},
  \citenamefont {Polloreno}, \citenamefont {Rawat}, \citenamefont {Ryan},
  \citenamefont {Renzas}, \citenamefont {Rubin}, \citenamefont {Russel},
  \citenamefont {Rust}, \citenamefont {Scarabelli}, \citenamefont
  {Selvanayagam}, \citenamefont {Sinclair}, \citenamefont {Smith},
  \citenamefont {Suska}, \citenamefont {To}, \citenamefont {Vahidpour},
  \citenamefont {Vodrahalli}, \citenamefont {Whyland}, \citenamefont {Yadav},
  \citenamefont {Zeng},\ and\ \citenamefont {Rigetti}}]{Reagoreaao3603}%
  \BibitemOpen
  \bibfield  {author} {\bibinfo {author} {\bibfnamefont {M.}~\bibnamefont
  {Reagor}}, \bibinfo {author} {\bibfnamefont {C.~B.}\ \bibnamefont {Osborn}},
  \bibinfo {author} {\bibfnamefont {N.}~\bibnamefont {Tezak}}, \bibinfo
  {author} {\bibfnamefont {A.}~\bibnamefont {Staley}}, \bibinfo {author}
  {\bibfnamefont {G.}~\bibnamefont {Prawiroatmodjo}}, \bibinfo {author}
  {\bibfnamefont {M.}~\bibnamefont {Scheer}}, \bibinfo {author} {\bibfnamefont
  {N.}~\bibnamefont {Alidoust}}, \bibinfo {author} {\bibfnamefont {E.~A.}\
  \bibnamefont {Sete}}, \bibinfo {author} {\bibfnamefont {N.}~\bibnamefont
  {Didier}}, \bibinfo {author} {\bibfnamefont {M.~P.}\ \bibnamefont
  {da~Silva}}, \bibinfo {author} {\bibfnamefont {E.}~\bibnamefont {Acala}},
  \bibinfo {author} {\bibfnamefont {J.}~\bibnamefont {Angeles}}, \bibinfo
  {author} {\bibfnamefont {A.}~\bibnamefont {Bestwick}}, \bibinfo {author}
  {\bibfnamefont {M.}~\bibnamefont {Block}}, \bibinfo {author} {\bibfnamefont
  {B.}~\bibnamefont {Bloom}}, \bibinfo {author} {\bibfnamefont
  {A.}~\bibnamefont {Bradley}}, \bibinfo {author} {\bibfnamefont
  {C.}~\bibnamefont {Bui}}, \bibinfo {author} {\bibfnamefont {S.}~\bibnamefont
  {Caldwell}}, \bibinfo {author} {\bibfnamefont {L.}~\bibnamefont {Capelluto}},
  \bibinfo {author} {\bibfnamefont {R.}~\bibnamefont {Chilcott}}, \bibinfo
  {author} {\bibfnamefont {J.}~\bibnamefont {Cordova}}, \bibinfo {author}
  {\bibfnamefont {G.}~\bibnamefont {Crossman}}, \bibinfo {author}
  {\bibfnamefont {M.}~\bibnamefont {Curtis}}, \bibinfo {author} {\bibfnamefont
  {S.}~\bibnamefont {Deshpande}}, \bibinfo {author} {\bibfnamefont
  {T.}~\bibnamefont {El~Bouayadi}}, \bibinfo {author} {\bibfnamefont
  {D.}~\bibnamefont {Girshovich}}, \bibinfo {author} {\bibfnamefont
  {S.}~\bibnamefont {Hong}}, \bibinfo {author} {\bibfnamefont {A.}~\bibnamefont
  {Hudson}}, \bibinfo {author} {\bibfnamefont {P.}~\bibnamefont {Karalekas}},
  \bibinfo {author} {\bibfnamefont {K.}~\bibnamefont {Kuang}}, \bibinfo
  {author} {\bibfnamefont {M.}~\bibnamefont {Lenihan}}, \bibinfo {author}
  {\bibfnamefont {R.}~\bibnamefont {Manenti}}, \bibinfo {author} {\bibfnamefont
  {T.}~\bibnamefont {Manning}}, \bibinfo {author} {\bibfnamefont
  {J.}~\bibnamefont {Marshall}}, \bibinfo {author} {\bibfnamefont
  {Y.}~\bibnamefont {Mohan}}, \bibinfo {author} {\bibfnamefont
  {W.}~\bibnamefont {O{\textquoteright}Brien}}, \bibinfo {author}
  {\bibfnamefont {J.}~\bibnamefont {Otterbach}}, \bibinfo {author}
  {\bibfnamefont {A.}~\bibnamefont {Papageorge}}, \bibinfo {author}
  {\bibfnamefont {J.-P.}\ \bibnamefont {Paquette}}, \bibinfo {author}
  {\bibfnamefont {M.}~\bibnamefont {Pelstring}}, \bibinfo {author}
  {\bibfnamefont {A.}~\bibnamefont {Polloreno}}, \bibinfo {author}
  {\bibfnamefont {V.}~\bibnamefont {Rawat}}, \bibinfo {author} {\bibfnamefont
  {C.~A.}\ \bibnamefont {Ryan}}, \bibinfo {author} {\bibfnamefont
  {R.}~\bibnamefont {Renzas}}, \bibinfo {author} {\bibfnamefont
  {N.}~\bibnamefont {Rubin}}, \bibinfo {author} {\bibfnamefont
  {D.}~\bibnamefont {Russel}}, \bibinfo {author} {\bibfnamefont
  {M.}~\bibnamefont {Rust}}, \bibinfo {author} {\bibfnamefont {D.}~\bibnamefont
  {Scarabelli}}, \bibinfo {author} {\bibfnamefont {M.}~\bibnamefont
  {Selvanayagam}}, \bibinfo {author} {\bibfnamefont {R.}~\bibnamefont
  {Sinclair}}, \bibinfo {author} {\bibfnamefont {R.}~\bibnamefont {Smith}},
  \bibinfo {author} {\bibfnamefont {M.}~\bibnamefont {Suska}}, \bibinfo
  {author} {\bibfnamefont {T.-W.}\ \bibnamefont {To}}, \bibinfo {author}
  {\bibfnamefont {M.}~\bibnamefont {Vahidpour}}, \bibinfo {author}
  {\bibfnamefont {N.}~\bibnamefont {Vodrahalli}}, \bibinfo {author}
  {\bibfnamefont {T.}~\bibnamefont {Whyland}}, \bibinfo {author} {\bibfnamefont
  {K.}~\bibnamefont {Yadav}}, \bibinfo {author} {\bibfnamefont
  {W.}~\bibnamefont {Zeng}}, \ and\ \bibinfo {author} {\bibfnamefont {C.~T.}\
  \bibnamefont {Rigetti}},\ }\href@noop {} {\bibfield  {journal} {\bibinfo
  {journal} {Sci. Adv.}\ }\textbf {\bibinfo {volume} {4}} (\bibinfo {year}
  {2018}{\natexlab{a}})}\BibitemShut {NoStop}%
\bibitem [{\citenamefont {Sirois}\ \emph {et~al.}(2015)\citenamefont {Sirois},
  \citenamefont {Castellanos-Beltran}, \citenamefont {DeFeo}, \citenamefont
  {Ranzani}, \citenamefont {Lecocq}, \citenamefont {Simmonds}, \citenamefont
  {Teufel},\ and\ \citenamefont {Aumentado}}]{doi:10.1063/1.4919759}%
  \BibitemOpen
  \bibfield  {author} {\bibinfo {author} {\bibfnamefont {A.~J.}\ \bibnamefont
  {Sirois}}, \bibinfo {author} {\bibfnamefont {M.~A.}\ \bibnamefont
  {Castellanos-Beltran}}, \bibinfo {author} {\bibfnamefont {M.~P.}\
  \bibnamefont {DeFeo}}, \bibinfo {author} {\bibfnamefont {L.}~\bibnamefont
  {Ranzani}}, \bibinfo {author} {\bibfnamefont {F.}~\bibnamefont {Lecocq}},
  \bibinfo {author} {\bibfnamefont {R.~W.}\ \bibnamefont {Simmonds}}, \bibinfo
  {author} {\bibfnamefont {J.~D.}\ \bibnamefont {Teufel}}, \ and\ \bibinfo
  {author} {\bibfnamefont {J.}~\bibnamefont {Aumentado}},\ }\href@noop {}
  {\bibfield  {journal} {\bibinfo  {journal} {Appl. Phys. Lett.}\ }\textbf
  {\bibinfo {volume} {106}},\ \bibinfo {pages} {172603} (\bibinfo {year}
  {2015})}\BibitemShut {NoStop}%
\bibitem [{\citenamefont {DiCarlo}\ \emph {et~al.}(2009)\citenamefont
  {DiCarlo}, \citenamefont {Chow}, \citenamefont {Gambetta}, \citenamefont
  {Bishop}, \citenamefont {Johnson}, \citenamefont {Schuster}, \citenamefont
  {Majer}, \citenamefont {Blais}, \citenamefont {Frunzio}, \citenamefont
  {Girvin},\ and\ \citenamefont {Schoelkopf}}]{DiCarlo2009}%
  \BibitemOpen
  \bibfield  {author} {\bibinfo {author} {\bibfnamefont {L.}~\bibnamefont
  {DiCarlo}}, \bibinfo {author} {\bibfnamefont {J.~M.}\ \bibnamefont {Chow}},
  \bibinfo {author} {\bibfnamefont {J.~M.}\ \bibnamefont {Gambetta}}, \bibinfo
  {author} {\bibfnamefont {L.~S.}\ \bibnamefont {Bishop}}, \bibinfo {author}
  {\bibfnamefont {B.~R.}\ \bibnamefont {Johnson}}, \bibinfo {author}
  {\bibfnamefont {D.~I.}\ \bibnamefont {Schuster}}, \bibinfo {author}
  {\bibfnamefont {J.}~\bibnamefont {Majer}}, \bibinfo {author} {\bibfnamefont
  {A.}~\bibnamefont {Blais}}, \bibinfo {author} {\bibfnamefont
  {L.}~\bibnamefont {Frunzio}}, \bibinfo {author} {\bibfnamefont {S.~M.}\
  \bibnamefont {Girvin}}, \ and\ \bibinfo {author} {\bibfnamefont {R.~J.}\
  \bibnamefont {Schoelkopf}},\ }\href@noop {} {\bibfield  {journal} {\bibinfo
  {journal} {Nature}\ }\textbf {\bibinfo {volume} {460}},\ \bibinfo {pages}
  {240} (\bibinfo {year} {2009})}\BibitemShut {NoStop}%
\bibitem [{\citenamefont {Lu}\ \emph {et~al.}(2017)\citenamefont {Lu},
  \citenamefont {Chakram}, \citenamefont {Leung}, \citenamefont {Earnest},
  \citenamefont {Naik}, \citenamefont {Huang}, \citenamefont {Groszkowski},
  \citenamefont {Kapit}, \citenamefont {Koch},\ and\ \citenamefont
  {Schuster}}]{Lu2017}%
  \BibitemOpen
  \bibfield  {author} {\bibinfo {author} {\bibfnamefont {Y.}~\bibnamefont
  {Lu}}, \bibinfo {author} {\bibfnamefont {S.}~\bibnamefont {Chakram}},
  \bibinfo {author} {\bibfnamefont {N.}~\bibnamefont {Leung}}, \bibinfo
  {author} {\bibfnamefont {N.}~\bibnamefont {Earnest}}, \bibinfo {author}
  {\bibfnamefont {R.~K.}\ \bibnamefont {Naik}}, \bibinfo {author}
  {\bibfnamefont {Z.}~\bibnamefont {Huang}}, \bibinfo {author} {\bibfnamefont
  {P.}~\bibnamefont {Groszkowski}}, \bibinfo {author} {\bibfnamefont
  {E.}~\bibnamefont {Kapit}}, \bibinfo {author} {\bibfnamefont
  {J.}~\bibnamefont {Koch}}, \ and\ \bibinfo {author} {\bibfnamefont {D.~I.}\
  \bibnamefont {Schuster}},\ }\href@noop {} {\bibfield  {journal} {\bibinfo
  {journal} {Phys. Rev. Lett.}\ }\textbf {\bibinfo {volume} {119}},\ \bibinfo
  {pages} {150502} (\bibinfo {year} {2017})}\BibitemShut {NoStop}%
\bibitem [{\citenamefont {Theis}\ and\ \citenamefont
  {Wilhelm}(2017)}]{Theis2017}%
  \BibitemOpen
  \bibfield  {author} {\bibinfo {author} {\bibfnamefont {L.~S.}\ \bibnamefont
  {Theis}}\ and\ \bibinfo {author} {\bibfnamefont {F.~K.}\ \bibnamefont
  {Wilhelm}},\ }\href@noop {} {\bibfield  {journal} {\bibinfo  {journal} {Phys.
  Rev. A}\ }\textbf {\bibinfo {volume} {95}},\ \bibinfo {pages} {022314}
  (\bibinfo {year} {2017})}\BibitemShut {NoStop}%
\bibitem [{\citenamefont {Koch}\ \emph {et~al.}(2007)\citenamefont {Koch},
  \citenamefont {Yu}, \citenamefont {Gambetta}, \citenamefont {Houck},
  \citenamefont {Schuster}, \citenamefont {Majer}, \citenamefont {Blais},
  \citenamefont {Devoret}, \citenamefont {Girvin},\ and\ \citenamefont
  {Schoelkopf}}]{Koch2007}%
  \BibitemOpen
  \bibfield  {author} {\bibinfo {author} {\bibfnamefont {J.}~\bibnamefont
  {Koch}}, \bibinfo {author} {\bibfnamefont {T.~M.}\ \bibnamefont {Yu}},
  \bibinfo {author} {\bibfnamefont {J.}~\bibnamefont {Gambetta}}, \bibinfo
  {author} {\bibfnamefont {A.~A.}\ \bibnamefont {Houck}}, \bibinfo {author}
  {\bibfnamefont {D.~I.}\ \bibnamefont {Schuster}}, \bibinfo {author}
  {\bibfnamefont {J.}~\bibnamefont {Majer}}, \bibinfo {author} {\bibfnamefont
  {A.}~\bibnamefont {Blais}}, \bibinfo {author} {\bibfnamefont {M.~H.}\
  \bibnamefont {Devoret}}, \bibinfo {author} {\bibfnamefont {S.~M.}\
  \bibnamefont {Girvin}}, \ and\ \bibinfo {author} {\bibfnamefont {R.~J.}\
  \bibnamefont {Schoelkopf}},\ }\href@noop {} {\bibfield  {journal} {\bibinfo
  {journal} {Phys. Rev. A}\ }\textbf {\bibinfo {volume} {76}},\ \bibinfo
  {pages} {042319} (\bibinfo {year} {2007})}\BibitemShut {NoStop}%
\bibitem [{\citenamefont {Johansson}\ \emph {et~al.}(2013)\citenamefont
  {Johansson}, \citenamefont {Nation},\ and\ \citenamefont
  {Nori}}]{JOHANSSON20131234}%
  \BibitemOpen
  \bibfield  {author} {\bibinfo {author} {\bibfnamefont {J.}~\bibnamefont
  {Johansson}}, \bibinfo {author} {\bibfnamefont {P.}~\bibnamefont {Nation}}, \
  and\ \bibinfo {author} {\bibfnamefont {F.}~\bibnamefont {Nori}},\ }\href@noop
  {} {\bibfield  {journal} {\bibinfo  {journal} {Comput. Phys. Commun.}\
  }\textbf {\bibinfo {volume} {184}},\ \bibinfo {pages} {1234 } (\bibinfo
  {year} {2013})}\BibitemShut {NoStop}%
\bibitem [{\citenamefont {Moll}\ \emph {et~al.}(2016)\citenamefont {Moll},
  \citenamefont {Fuhrer}, \citenamefont {Staar},\ and\ \citenamefont
  {Tavernelli}}]{Moll2016}%
  \BibitemOpen
  \bibfield  {author} {\bibinfo {author} {\bibfnamefont {N.}~\bibnamefont
  {Moll}}, \bibinfo {author} {\bibfnamefont {A.}~\bibnamefont {Fuhrer}},
  \bibinfo {author} {\bibfnamefont {P.}~\bibnamefont {Staar}}, \ and\ \bibinfo
  {author} {\bibfnamefont {I.}~\bibnamefont {Tavernelli}},\ }\href@noop {}
  {\bibfield  {journal} {\bibinfo  {journal} {J. Phys. A}\ }\textbf {\bibinfo
  {volume} {49}},\ \bibinfo {pages} {295301} (\bibinfo {year}
  {2016})}\BibitemShut {NoStop}%
\bibitem [{\citenamefont {Albash}\ and\ \citenamefont
  {Lidar}(2015)}]{Albash2015}%
  \BibitemOpen
  \bibfield  {author} {\bibinfo {author} {\bibfnamefont {T.}~\bibnamefont
  {Albash}}\ and\ \bibinfo {author} {\bibfnamefont {D.~A.}\ \bibnamefont
  {Lidar}},\ }\href@noop {} {\bibfield  {journal} {\bibinfo  {journal} {Phys.
  Rev. A}\ }\textbf {\bibinfo {volume} {91}},\ \bibinfo {pages} {062320}
  (\bibinfo {year} {2015})}\BibitemShut {NoStop}%
\bibitem [{\citenamefont {Malossi}\ \emph {et~al.}(2013)\citenamefont
  {Malossi}, \citenamefont {Bason}, \citenamefont {Viteau}, \citenamefont
  {Arimondo}, \citenamefont {Mannella}, \citenamefont {Morsch},\ and\
  \citenamefont {Ciampini}}]{PhysRevA.87.012116}%
  \BibitemOpen
  \bibfield  {author} {\bibinfo {author} {\bibfnamefont {N.}~\bibnamefont
  {Malossi}}, \bibinfo {author} {\bibfnamefont {M.~G.}\ \bibnamefont {Bason}},
  \bibinfo {author} {\bibfnamefont {M.}~\bibnamefont {Viteau}}, \bibinfo
  {author} {\bibfnamefont {E.}~\bibnamefont {Arimondo}}, \bibinfo {author}
  {\bibfnamefont {R.}~\bibnamefont {Mannella}}, \bibinfo {author}
  {\bibfnamefont {O.}~\bibnamefont {Morsch}}, \ and\ \bibinfo {author}
  {\bibfnamefont {D.}~\bibnamefont {Ciampini}},\ }\href@noop {} {\bibfield
  {journal} {\bibinfo  {journal} {Phys. Rev. A}\ }\textbf {\bibinfo {volume}
  {87}},\ \bibinfo {pages} {012116} (\bibinfo {year} {2013})}\BibitemShut
  {NoStop}%
\bibitem [{\citenamefont {Kounalakis}\ \emph {et~al.}(2018)\citenamefont
  {Kounalakis}, \citenamefont {Dickel}, \citenamefont {Bruno}, \citenamefont
  {Langford},\ and\ \citenamefont {Steele}}]{Kounalakis2018}%
  \BibitemOpen
  \bibfield  {author} {\bibinfo {author} {\bibfnamefont {M.}~\bibnamefont
  {Kounalakis}}, \bibinfo {author} {\bibfnamefont {C.}~\bibnamefont {Dickel}},
  \bibinfo {author} {\bibfnamefont {A.}~\bibnamefont {Bruno}}, \bibinfo
  {author} {\bibfnamefont {N.~K.}\ \bibnamefont {Langford}}, \ and\ \bibinfo
  {author} {\bibfnamefont {G.~A.}\ \bibnamefont {Steele}},\ }\href@noop {}
  {\bibfield  {journal} {\bibinfo  {journal} {npj Quantum Inf.}\ }\textbf
  {\bibinfo {volume} {4}},\ \bibinfo {pages} {38} (\bibinfo {year}
  {2018})}\BibitemShut {NoStop}%
\bibitem [{\citenamefont {Reagor}\ \emph
  {et~al.}(2018{\natexlab{b}})\citenamefont {Reagor}, \citenamefont {Osborn},
  \citenamefont {Tezak}, \citenamefont {Staley}, \citenamefont
  {Prawiroatmodjo}, \citenamefont {Scheer}, \citenamefont {Alidoust},
  \citenamefont {Sete}, \citenamefont {Didier}, \citenamefont {da~Silva},
  \citenamefont {Acala}, \citenamefont {Angeles}, \citenamefont {Bestwick},
  \citenamefont {Block}, \citenamefont {Bloom}, \citenamefont {Bradley},
  \citenamefont {Bui}, \citenamefont {Caldwell}, \citenamefont {Capelluto},
  \citenamefont {Chilcott}, \citenamefont {Cordova}, \citenamefont {Crossman},
  \citenamefont {Curtis}, \citenamefont {Deshpande}, \citenamefont
  {El~Bouayadi}, \citenamefont {Girshovich}, \citenamefont {Hong},
  \citenamefont {Hudson}, \citenamefont {Karalekas}, \citenamefont {Kuang},
  \citenamefont {Lenihan}, \citenamefont {Manenti}, \citenamefont {Manning},
  \citenamefont {Marshall}, \citenamefont {Mohan}, \citenamefont
  {O{\textquoteright}Brien}, \citenamefont {Otterbach}, \citenamefont
  {Papageorge}, \citenamefont {Paquette}, \citenamefont {Pelstring},
  \citenamefont {Polloreno}, \citenamefont {Rawat}, \citenamefont {Ryan},
  \citenamefont {Renzas}, \citenamefont {Rubin}, \citenamefont {Russel},
  \citenamefont {Rust}, \citenamefont {Scarabelli}, \citenamefont
  {Selvanayagam}, \citenamefont {Sinclair}, \citenamefont {Smith},
  \citenamefont {Suska}, \citenamefont {To}, \citenamefont {Vahidpour},
  \citenamefont {Vodrahalli}, \citenamefont {Whyland}, \citenamefont {Yadav},
  \citenamefont {Zeng},\ and\ \citenamefont {Rigetti}}]{Reagor2018}%
  \BibitemOpen
  \bibfield  {author} {\bibinfo {author} {\bibfnamefont {M.}~\bibnamefont
  {Reagor}}, \bibinfo {author} {\bibfnamefont {C.~B.}\ \bibnamefont {Osborn}},
  \bibinfo {author} {\bibfnamefont {N.}~\bibnamefont {Tezak}}, \bibinfo
  {author} {\bibfnamefont {A.}~\bibnamefont {Staley}}, \bibinfo {author}
  {\bibfnamefont {G.}~\bibnamefont {Prawiroatmodjo}}, \bibinfo {author}
  {\bibfnamefont {M.}~\bibnamefont {Scheer}}, \bibinfo {author} {\bibfnamefont
  {N.}~\bibnamefont {Alidoust}}, \bibinfo {author} {\bibfnamefont {E.~A.}\
  \bibnamefont {Sete}}, \bibinfo {author} {\bibfnamefont {N.}~\bibnamefont
  {Didier}}, \bibinfo {author} {\bibfnamefont {M.~P.}\ \bibnamefont
  {da~Silva}}, \bibinfo {author} {\bibfnamefont {E.}~\bibnamefont {Acala}},
  \bibinfo {author} {\bibfnamefont {J.}~\bibnamefont {Angeles}}, \bibinfo
  {author} {\bibfnamefont {A.}~\bibnamefont {Bestwick}}, \bibinfo {author}
  {\bibfnamefont {M.}~\bibnamefont {Block}}, \bibinfo {author} {\bibfnamefont
  {B.}~\bibnamefont {Bloom}}, \bibinfo {author} {\bibfnamefont
  {A.}~\bibnamefont {Bradley}}, \bibinfo {author} {\bibfnamefont
  {C.}~\bibnamefont {Bui}}, \bibinfo {author} {\bibfnamefont {S.}~\bibnamefont
  {Caldwell}}, \bibinfo {author} {\bibfnamefont {L.}~\bibnamefont {Capelluto}},
  \bibinfo {author} {\bibfnamefont {R.}~\bibnamefont {Chilcott}}, \bibinfo
  {author} {\bibfnamefont {J.}~\bibnamefont {Cordova}}, \bibinfo {author}
  {\bibfnamefont {G.}~\bibnamefont {Crossman}}, \bibinfo {author}
  {\bibfnamefont {M.}~\bibnamefont {Curtis}}, \bibinfo {author} {\bibfnamefont
  {S.}~\bibnamefont {Deshpande}}, \bibinfo {author} {\bibfnamefont
  {T.}~\bibnamefont {El~Bouayadi}}, \bibinfo {author} {\bibfnamefont
  {D.}~\bibnamefont {Girshovich}}, \bibinfo {author} {\bibfnamefont
  {S.}~\bibnamefont {Hong}}, \bibinfo {author} {\bibfnamefont {A.}~\bibnamefont
  {Hudson}}, \bibinfo {author} {\bibfnamefont {P.}~\bibnamefont {Karalekas}},
  \bibinfo {author} {\bibfnamefont {K.}~\bibnamefont {Kuang}}, \bibinfo
  {author} {\bibfnamefont {M.}~\bibnamefont {Lenihan}}, \bibinfo {author}
  {\bibfnamefont {R.}~\bibnamefont {Manenti}}, \bibinfo {author} {\bibfnamefont
  {T.}~\bibnamefont {Manning}}, \bibinfo {author} {\bibfnamefont
  {J.}~\bibnamefont {Marshall}}, \bibinfo {author} {\bibfnamefont
  {Y.}~\bibnamefont {Mohan}}, \bibinfo {author} {\bibfnamefont
  {W.}~\bibnamefont {O{\textquoteright}Brien}}, \bibinfo {author}
  {\bibfnamefont {J.}~\bibnamefont {Otterbach}}, \bibinfo {author}
  {\bibfnamefont {A.}~\bibnamefont {Papageorge}}, \bibinfo {author}
  {\bibfnamefont {J.-P.}\ \bibnamefont {Paquette}}, \bibinfo {author}
  {\bibfnamefont {M.}~\bibnamefont {Pelstring}}, \bibinfo {author}
  {\bibfnamefont {A.}~\bibnamefont {Polloreno}}, \bibinfo {author}
  {\bibfnamefont {V.}~\bibnamefont {Rawat}}, \bibinfo {author} {\bibfnamefont
  {C.~A.}\ \bibnamefont {Ryan}}, \bibinfo {author} {\bibfnamefont
  {R.}~\bibnamefont {Renzas}}, \bibinfo {author} {\bibfnamefont
  {N.}~\bibnamefont {Rubin}}, \bibinfo {author} {\bibfnamefont
  {D.}~\bibnamefont {Russel}}, \bibinfo {author} {\bibfnamefont
  {M.}~\bibnamefont {Rust}}, \bibinfo {author} {\bibfnamefont {D.}~\bibnamefont
  {Scarabelli}}, \bibinfo {author} {\bibfnamefont {M.}~\bibnamefont
  {Selvanayagam}}, \bibinfo {author} {\bibfnamefont {R.}~\bibnamefont
  {Sinclair}}, \bibinfo {author} {\bibfnamefont {R.}~\bibnamefont {Smith}},
  \bibinfo {author} {\bibfnamefont {M.}~\bibnamefont {Suska}}, \bibinfo
  {author} {\bibfnamefont {T.-W.}\ \bibnamefont {To}}, \bibinfo {author}
  {\bibfnamefont {M.}~\bibnamefont {Vahidpour}}, \bibinfo {author}
  {\bibfnamefont {N.}~\bibnamefont {Vodrahalli}}, \bibinfo {author}
  {\bibfnamefont {T.}~\bibnamefont {Whyland}}, \bibinfo {author} {\bibfnamefont
  {K.}~\bibnamefont {Yadav}}, \bibinfo {author} {\bibfnamefont
  {W.}~\bibnamefont {Zeng}}, \ and\ \bibinfo {author} {\bibfnamefont {C.~T.}\
  \bibnamefont {Rigetti}},\ }\href@noop {} {\bibfield  {journal} {\bibinfo
  {journal} {Sci. Adv.}\ }\textbf {\bibinfo {volume} {4}} (\bibinfo {year}
  {2018}{\natexlab{b}})}\BibitemShut {NoStop}%
\end{thebibliography}%
\end{document}